\documentclass[pre,aps,twocolumn,showpacs,floatfix,10pt,superscriptaddress]{revtex4-2}
\usepackage{amsmath}
\usepackage{amsfonts}
\usepackage{amssymb}
\usepackage{graphicx}
\usepackage{bm}
\usepackage{color}
\usepackage{ulem}
 
 \usepackage{epsfig}
\usepackage{graphicx}
\usepackage{chngcntr}
\counterwithout{figure}{section}
\usepackage{rotating}
\usepackage{amssymb}
\usepackage{amsmath}
\usepackage{physics}
\usepackage{multirow}
\usepackage{float}
\usepackage{caption, subcaption}

\begin{document}
\title{Hysteresis and return point memory in the random field Blume Capel model}
\author{B.E. Aldrin}
\email{aldrin.be@niser.ac.in}
\affiliation{School of Physical Sciences, National Institute of Science Education and Research, Bhubaneswar, P.O. Jatni, Khurda, Odisha, India 752050}
\affiliation{Homi Bhabha National Institute, Training School Complex, Anushakti Nagar, Mumbai 400094, India}
\author{Abdul Khaleque}
\email{aktphys@gmail.com}
\affiliation{Department of Physics, Bidhan Chandra College, University of Calcutta, Kolkata, India}
\author{Sumedha}
\email{sumedha@niser.ac.in}
\affiliation{School of Physical Sciences, National Institute of Science Education and Research, Bhubaneswar, P.O. Jatni, Khurda, Odisha, India 752050}
\affiliation{Homi Bhabha National Institute, Training School Complex, Anushakti Nagar, Mumbai 400094, India}

\begin{abstract}
We study the zero temperature steady state of the random field Blume Capel model with spin-flip Glauber dynamics on a random regular graph. The magnetization $m$ as a function of the external field $H$ is observed to have double hysteresis loops with a return point memory. We also solve the model on a Bethe lattice in the approximation that the spin relaxation dynamics is abelian and find good agreement between simulations on random regular graphs and Bethe lattice calculations for negative values of $H$. 

\end{abstract}


\maketitle

\section{Introduction}

In experiments, it is often seen that a piece of magnet exposed to an increasing external magnetic field magnetises in a series of jumps or avalanches. This gives rise to a magnetization that lags behind the external magnetic field and shows a history dependence. This effect is known as hysteresis \cite{sethnabook}. The magnetic signal produced on changing the external field known as the Barkhausen noise, is highly reproducible \cite{noise}. Hysteresis is also seen in many other systems (see \cite{keim} for a recent review) like martensites \cite{ortin1}, DNA unzipping \cite{dna}, granular matter \cite{granular1,granular2}, output vs input in economics \cite{economy} and perception in psychology \cite{psychology}.

Systems with multiple metastable states with energy barriers higher than the energy of the thermal fluctuations typically exhibit hysteresis on the application of an external force. Spin systems with quenched disorder are examples of systems with multiple metastable states. Sethna et al \cite{sethna1}  studied the zero temperature random-field Ising model (RFIM)  and found that the model exhibits hysteresis and a return point memory (RPM). The zero temperature RFIM has been very useful in understanding the phenomenon of hysteresis and RPM. The model has hence been studied extensively and continues to be of interest\cite{noise,deepak,sanjib1,balog,vives,nandi,rosinberg,hayden,clemmer,ferre,
shukla1,shukla2,ramola,mijatovic}. RPM can be described as the system’s ability to return to a previous state when the direction of the applied magnetic field is reversed. This is important as it allows magnetic tapes to be rerecorded and gives shape memory to shape memory alloys \cite{keim}. The existence of RPM  for zero temperature RFIM steady state with spin-flip Glauber dynamics \cite{glauber} was proved using the no passing property \cite{sethna1,middleton}. 

The steady state magnetization for the RFIM with zero temperature Glauber spin-flip dynamics exhibits a single hysteresis loop as the external field is varied. In experiments, one often observes multiple hysteresis loops in systems like ferroelectrics \cite{doublehys1,doublehys2}, antiferroelectrics \cite{doublehys3} and martensites \cite{doublehys4}. A connection between multiple hysteresis and higher spin molecules was reported in experiments involving $Mn_{12}$ \cite{mulhys}. In most of these systems, like the thermoelastic martensites, RPM is also seen in experiments \cite{ortin1}. Numerical studies involving higher spins and continuous spins with random fields have also reported multiple hysteresis loops  \cite{ortin2,akinci,shukla}. 

In this paper we study the hysteresis response of the zero temperature spin-$1$ random field Blume Capel model (RFBCM) and find that the magnetization exhibits double hysteresis (see Fig. \ref{hystsim}) with RPM. The model is defined by the Hamiltonian:
\begin{eqnarray}
 H&=& -\sum_{<ij>} s_i s_j-\sum_{i}h_i s_i- H \sum_{i} s_i+\Delta \sum_{i} s_i^2
\end{eqnarray}
Here $s_i$ are the spin-$1$ variables  that take values $\pm1$ and $0$. The index $i$ and $j$ run over all the lattice sites. $\Delta$ is the crystal field strength : a positive value of $\Delta$ favours the $0$ 
spin state, while a negative value favours the $\pm 1$ spin states. 
$H$ is the external magnetic field and $h_i$ is the quenched local 
random field, drawn from a  continuous Gaussian probability 
distribution $\phi (h)= \frac{1}{ \sqrt{2 \pi \sigma^2}} e^{-h^2/{2 \sigma^2}}$, with $\sigma$ the standard deviation of the 
distribution. At $T=0$ in the absence of the random field, the model has a first order transition from an ordered state to a disordered state as a function of $\Delta$. In the presence of the random Gaussian field the equilibrium phase diagram of the model on a fully connected graph has been studied recently \cite{soheli}. The phase diagram in the $\Delta-\sigma$ plane has a line of critical transition ending at a tricritical point. The zero temperature phase diagram of the model on random regular graphs and finite dimensional lattices has not been studied and numerical studies similar to that of RFIM will be useful for this model as well \cite{fytas}.

We study the $T=0$ RFBCM numerically using the zero temperature Glauber spin-flip dynamics on a random regular graph. At $T=0$, the Glauber spin-flip dynamics lacks detailed balance and we get a non-equilibirum steady state in which the magnetization shows hysteresis. It is useful to study models on a random regular graph as they take correlations between neighbouring sites into account. The nature of the steady state can be different on a random regular graph than on a fully connected graph, in the presence of disorder. For example, zero temperature RFIM on a fully connected graph shows hysteresis only for $\sigma$ less than a critical value, whereas in the case of random regular graphs, hysteresis is seen for all values of $\sigma$ \cite{sethna1,deepak}.

We find that even though the dynamics is not abelian \cite{footnote} for RFBCM, zero temperature steady state of the model exhibits RPM on a random regular graph. We solve the model on a Bethe lattice by extending the method used to study the RFIM \cite{deepak}. The hysteresis plots obtained on the Bethe lattice match with the plots for the same coordination number on a random regular graph for negative values of $H$ and do not match for positive values of $H$. Since the hysteresis loops on a random regular graph for RFBCM for positive and negative $H$ are identical on reversing the sign, our Bethe lattice calculations provide an approximate solution of the problem on a random regular graph. We also find that the magnetization undergoes a transition from first order hysteresis to second order hysteresis for RFBCM on random regular graphs with coordination number $ \ge 3$ at a value of $\sigma$ that changes with $\Delta$. This is in contrast to the behaviour of RFIM where first order hysteresis loops are seen only for coordination number $ \ge 4$ \cite{deepak}.

The plan of the paper is as follows: In Sec. \ref{srrg} we study hysteresis on random regular graphs and show the RPM property of the hysteresis curves. In Sec. \ref{sbl} we solve the model on the Bethe lattice and compare these results with the simulation results on a random regular graph. We summarise in Sec. \ref{sc}.

\section{Hysteresis on a random regular graph}
\label{srrg}
We study the zero temperature non-equilibrium steady state reached when the magnetic field is increased adiabatically from minus infinity to plus infinity on a random regular graph of coordination number $z$. The quenched local random field $h_i$ at each site is drawn from a Gaussian distribution $\phi(h_i)$ with standard deviation $\sigma$. To study hysteresis start with an initial state in which the external field $H$ is large and negative, such that the steady state of the system has all spins in $-1$ state. The field is increased in steps of $\delta H=0.01$. After each increment in $H$, the system relaxes via zero temperature single spin-flip Glauber dynamics till a steady state is reached (see Appendix \ref{appA} for details). In zero temperature Glauber spin-flip dynamics, a spin changes its state if the change in energy ($\delta_E$) due to the spin-flip is negative. Since there are two possible states to which a spin can flip, the one which lowers the energy more is chosen. The whole lattice is updated till there are no flippable spins. The change in energy depends on the initial and final spin state. There are three possibilities for a spin ($s$) at site $i$ to flip: 
\begin{itemize}
\item $s=\pm 1$ to $\mp 1$ : $\delta_E = 2 s (\sum_{j \in \mathcal{N}(i)} s_j + h_i  + H)$
\item $s=\pm 1$ to $0$ : $\delta_E = s \left(\sum_{j \in \mathcal{N}(i)} s_j + h_i  + H\right) -  \Delta$
\item $0$ to $s=\pm 1$ : $\delta_E = -s (\sum_{j \in \mathcal{N}(i)} s_j + h_i  + H) + \Delta$
\end{itemize}
here $\mathcal{N}(i)$ denotes the set of neighbours of the site $i$. 

We considered many different network configurations and found that  for large system sizes the change in magnetization from one realization of disorder to the other is not noticeable (see Appendix \ref{appA}). Simulations were done with $N=10^5$, $N$ being the system size. The magnetization is simply the sum of all spin values in the steady state divided by the system size $N$. 
\subsection{Hysteresis} 
\begin{figure}
 \includegraphics[width=0.9\columnwidth]{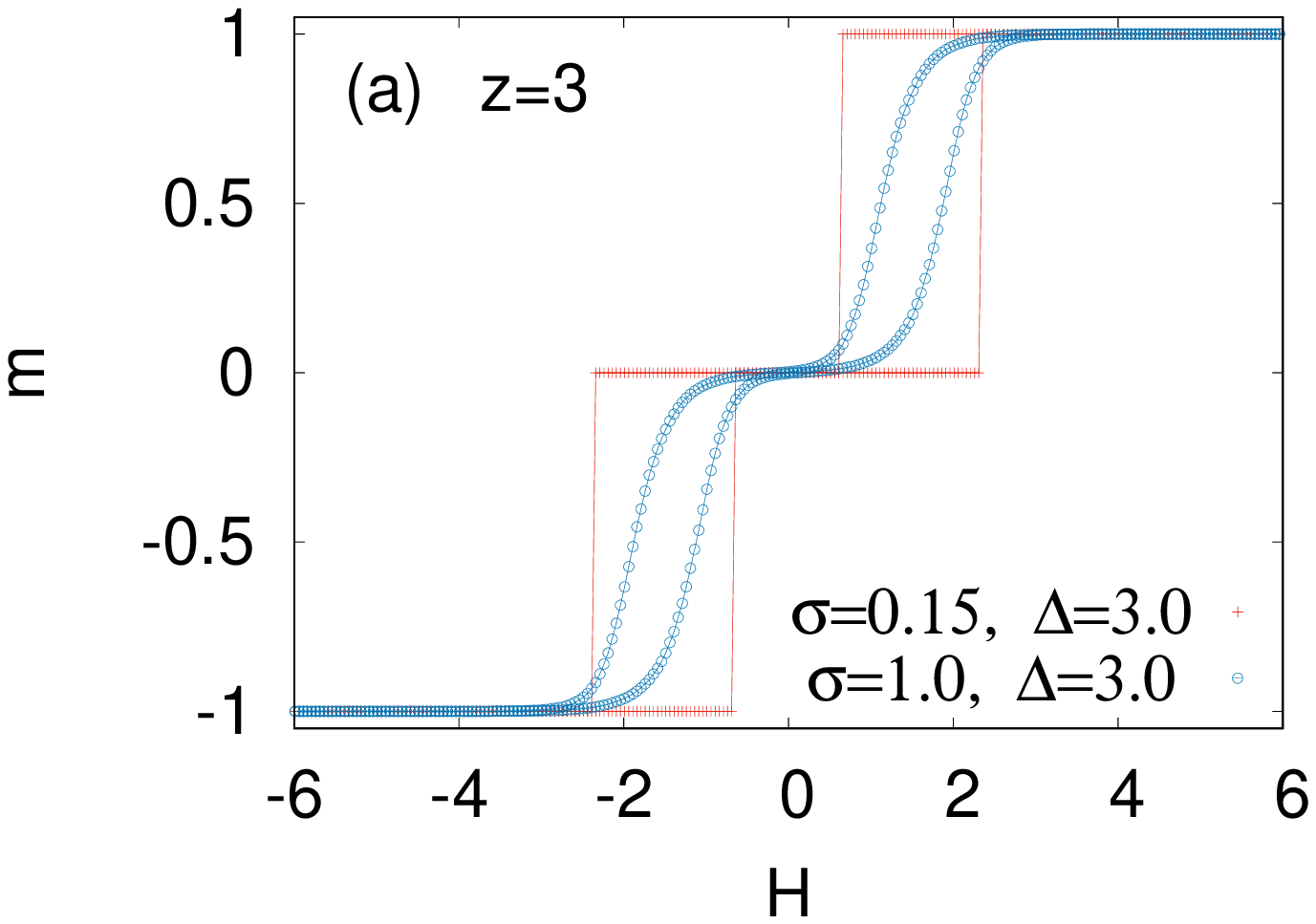}
  \includegraphics[width=0.9\columnwidth]{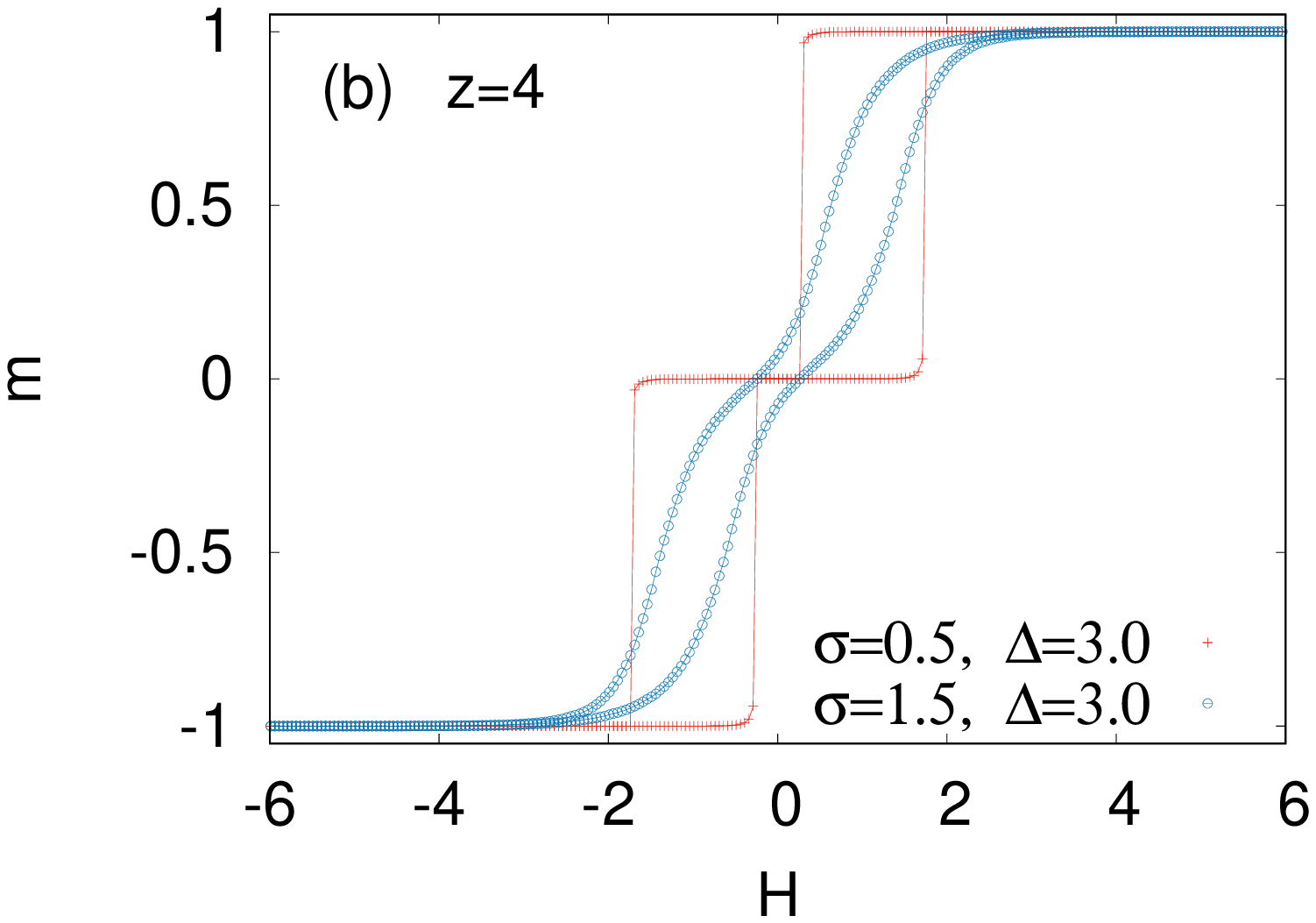}
   \caption{Hysteresis response of magnetization ($m$) for random regular graph with coordination number $z=3$ and $z=4$ is plotted for two different $\sigma$ values as a function of external field ($H$) for a typical realization of the disorder. In Fig. 1a, the red (light gray) line of plusses and the blue (dark gray) line of circles correspond to $\sigma = 0.15$ and $1.0$ respectively. In Fig. 1b, the red (light gray) and the blue (dark gray) lines correspond to $\sigma = 0.5$ and $1.5$ respectively. In both cases the hysteresis loop shows discontinuity for small $\sigma$ and transitions to continuous hysteresis loop for large $\sigma$. The value of $\sigma$ at which the transition occurs increases with $z$. For $z=3$ and $\Delta=3$, we find $\sigma_c = 0.40(5)$ and for $z=4$ and $\Delta=3$ we find $\sigma_c = 0.9(2)$.}
\label{hystsim}
\end{figure} 
The magnetization in the steady state at the same value of $H$ is different when the system is evolved starting with a very large negative external field  with all spins down than when the systems is evolved starting with a very large positive external field with all spins up.


We find that the RFBCM shows two jumps in the magnetization and hence double hysteresis for positive values of $\Delta$. In Fig. \ref{hystsim} the hysteresis curves for $z=3$ and $z=4$ for different values of $\Delta$ and $\sigma$ are shown. The magnetization jumps discontinuously for smaller $\sigma$ and  smoothens as $\sigma$ is increased. 
The magnetization undergoes a transition from being discontinuous to continuous as $\sigma$ is increased for all $z \ge 3$. 

For $\Delta \le 0$, the $\pm 1$ spin states are favoured over the $0$ spin state and the RFBCM has a behaviour similar to the RFIM with 
single hysteresis loops. In Appendix \ref{appB} we give the magnetization plots of different values for $z$, $\Delta$ and $\sigma$.

\subsection{Return-Point Memory (RPM)}
\begin{figure}
\includegraphics[width=0.9\columnwidth]{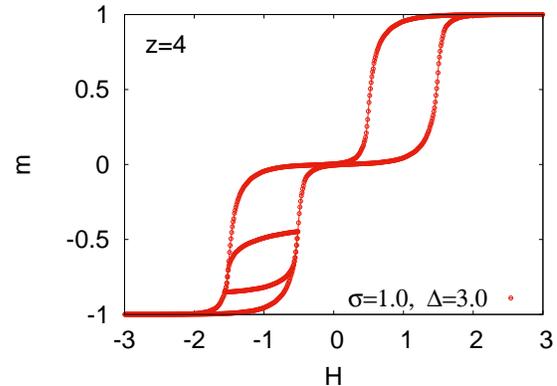}
   \caption{{Double hysteresis loops for zero temperature RFBCM between magnetization ($m$) and external field ($H$), showing RPM. The two hysteresis loops are identical and RPM exists between any two different values of $H$.}}
\label{RPM}
\end{figure}

On changing the external field from a value $H_1$ at time $t=0$ to $H_2$ at time $t=t_1$ and backtracking from $H_2$ to $H_1$, if the magnetization returns to its original value independent of the time $t_1$ and details of the evolution of the external field, then the system has RPM. We find magnetization in the case of RFBCM exhibits RPM  for all values of $\Delta$ and $\sigma$. For $z=4, \Delta=3$ and $\sigma=1$ it is shown in Fig. \ref{RPM}.

RPM is widely observed in experiments. In the case of RFIM, the presence of RPM was proven using the existence of  no passing property \cite{middleton}. No passing property for spin systems implies that if we take two configurations $C^{\alpha}$ and $C^{\beta}$ such that the spins in the two configurations are partially ordered ( for all the spins in the two configurations  $s_i^{\alpha} \ge s_i^{\beta}$ ), then the partial ordering is preserved under the application of the external field as long as the field $H^{\alpha}(t)$ applied to $C^{\alpha}$ is always greater than or equal to the field $H^{\beta}(t)$ applied to $C^{\beta}$. The no passing property implies abelian dynamics for zero temperature RFIM. The abelian dynamics for RFIM means that if we start from a configuration with all spins in the $-1$ state and external field $-\infty$ and increase the field to a finite value $H$ in one step then the  
order in which unstable spins are relaxed does not matter \cite{deepak,sanjib}.

No passing property holds also for the zero temperature RFBCM. The dynamics is not abelian for RFBCM i.e, the order of flipping of the unstable spins now can change the configuration attained. Although if we combine the $0$ and $+1$ states into one state $X$, we can think of the RFBCM as an effective two state system with abelian dynamics. In Section \ref{sbl}, we make use of this observation to find the magnetization of the model on a Bethe lattice.

\subsection{Coercive field and remanent magnetization}

\begin{figure}
\includegraphics[width=0.9\columnwidth]{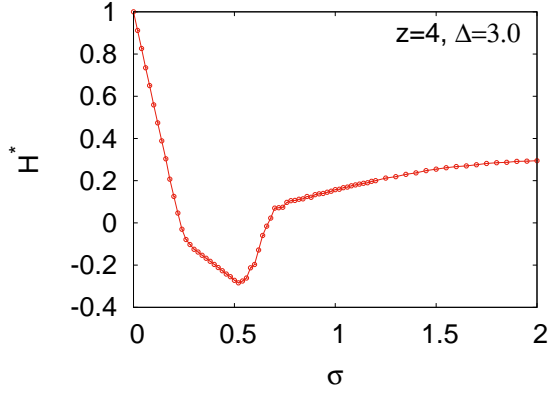}
   \caption{{Coercive field ($H^*$) for $z=4$, $\Delta=3$} as a function of $\sigma$.}
\label{cf}
\end{figure}

The coercive field $H^*$ is defined as the value of $H$ beyond which the negative $m$ solution disappears \cite{glauber}. Thus, in simulations while evolving the system, increasing the external field from a large negative value,  the coercive field is taken to be the field $H^*$ for which the magnetization $m$ of the system becomes non-negative for the first time. As shown in Fig. \ref{cf}, at small $\sigma$, $H^*$ decays rapidly. This is because the hysteresis loop is first order in this region and moves left on increasing $\sigma$ as more and more spins take the $0$ value. However as $\sigma$ increases further, $H^*$ increases with increasing $\sigma$. This change in 
slope is observed to occur before $\sigma_c$.

\begin{figure}
\includegraphics[width=1\columnwidth]{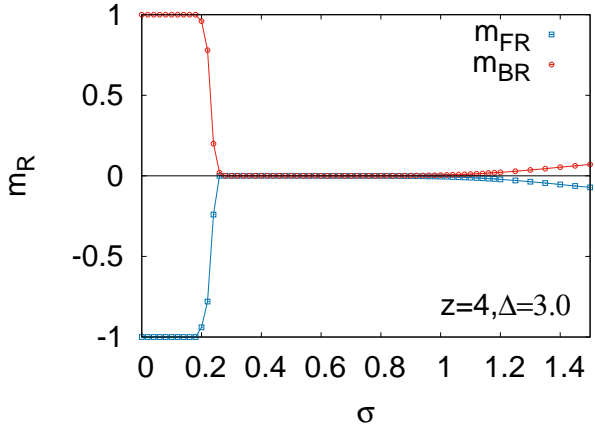}
   \caption{{Remanent magnetization ($m_R$) plot for $z=4$, $\Delta=3$. Here the blue (lower) line, $m_{FR}$ and  the red (upper) line, $m_{BR}$ denote the remanent magnetization $m_R$ curves obtained from forward and backward increments in H.}}
\label{rm}
\end{figure}

Another useful property is the remanent magnetization ($m_R$) which is the steady state value of the magnetization when starting from the initial state with all spins up or down, the external field $H$ is made zero \cite{glauber}. For RFBCM at  $\sigma=0$ the value of $m_R$ is $z-\Delta$. On increasing  $\sigma$, it first decreases as the first order hysteresis loop moves towards the left along the $H$ axis. It goes to zero  $0$ at a value of $\sigma$ which is well before the transition from the first order hysteresis to the continuous hysteresis. It continues to be zero for $\sigma<\sigma_c$ and increases with $\sigma$ beyond that (see Fig. \ref{rm}). 
\section{Model on the Bethe lattice}
\label{sbl}
For large system sizes, the loops in the random regular graph can be ignored after which it is like a Bethe lattice of the same coordination number. A Bethe lattice is the interior of a Cayley tree, such that the surface of the tree does not influence the statistical averages. The zero temperature hysteresis in RFIM was solved making use of the abelian property of the dynamics and it was shown that there was an excellent match with the simulations on a random regular graph \cite{deepak}. 
\begin{figure}
\includegraphics[width=\columnwidth]{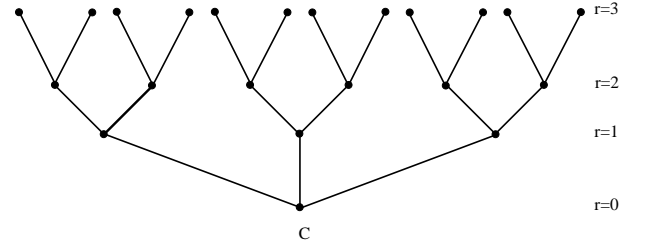}
   \caption{Cayley tree with coordination number $z=3$ and height $4$.}
\label{cayley}
\end{figure}

In the case of the RFBCM that we study here, since the spins can flip to either a $0$ state or $+1$ state on increasing $H$, the order of the spin-flip can change the configuration achieved.
The model exhibits RPM in spite of the absence of the abelian property of the dynamics. The abelian property still holds if we consider a two state system with $-1$ spin state and another state $X$, which can either be $0$ or $+1$. We hence expect the dynamics to be almost abelian for large negative $H$. We calculate the magnetization for the model on a Bethe lattice of coordination number $z$ in this section under this assumption and find good agreement with the simulations. 

Consider a Cayley tree where all non-boundary sites have coordination number $z$ and the boundary sites have coordination number $1$. Different levels of the tree are labelled by $r$ 
with $r=0$ for the root and $r=n$ for the boundary sites. Fig. \ref{cayley} shows a Cayley tree with $n=4$ levels for $z=3$. 
Start with $H=-\infty$, with all spins in the $-1$ state. Now increase the field to $H$ in a single step and relax the spins starting with the spins in the level $r=n$, keeping all the spins in other levels frozen in $-1$ state. Next relax the spins in level $r=n-1$, keeping the spins in levels below $n-1$ fixed in $-1$ state. Continue this process till all the spins on the tree are relaxed.

Since a spin can flip either to a $0$ state or a $+1$ state, we define probability $P_r$ ($Q_r$) to be the probability that a spin at level $r$ will flip to $+1$ ($0$) state given that the spins in levels $r+1$ and above have already been relaxed and the spins in 
levels $r-1$ and below are all in $-1$ state. The spin-flip probabilities of different spins in the same level are independent of each other and the recursion relations for $P_r$ and $Q_r$ are
\begin{gather}
 P_r= \sum_{s_0=0}^{(z-1)} \sum_{s_1=0}^{(z-1)-s_0} \frac{(z-1)!}{s_0!s_1!((z-1)-s_1-s_0)!} 
 Q_{r+1}^{s_0} \nonumber\\
  P_{r+1}^{s_1} (1-Q_{r+1}-P_{r+1})^{(z-1)-s_0-s_1} p_{s_0 s_1}
 \label{couple1}
\end{gather}
\begin{gather}
 Q_r= \sum_{s_0=0}^{(z-1)} \sum_{s_1=0}^{(z-1)-s_0} \frac{(z-1)!}{s_0!s_1!((z-1)-s_1-s_0)!} 
 Q_{r+1}^{s_0}\nonumber\\
   P_{r+1}^{s_1} (1-Q_{r+1}-P_{r+1})^{(z-1)-s_0-s_1} q_{s_0 s_1}
 \label{couple2}
\end{gather}
where $p_{s_0 s_1}$($q_{s_0 s_1}$) is the probability that a $-1$ spin  with $s_1$ descendants in $+1$ state and $s_0$ in $0$ state will flip to $+1$($0$) state. They are given by
\begin{eqnarray}
p_{s_0 s_1}=\int_{-H+(z-2s_1-s_0)+\Delta}^{\infty} \phi(h_i) dh_i.
\end{eqnarray}
\begin{eqnarray}
q_{s_0 s_1}=\int_{-H+(z-2s_1-s_0)-\Delta}^{-H+(z-2s_1-s_0)+\Delta} \phi(h_i) dh_i.
\end{eqnarray}
As $n \rightarrow \infty$, $P_r$ and $Q_r$ take their fixed point values $P^*$ and $Q^*$ respectively, which have recursions given by
\begin{gather}
 P^*= \sum_{s_0=0}^{(z-1)} \sum_{s_1=0}^{(z-1)-s_0} \frac{(z-1)!}{s_0!s_1!((z-1)-s_1-s_0)!}
 (Q^*)^{s_0} \nonumber\\
  (P^*)^{s_1} (1-Q^*-P^*)^{(z-1)-s_0-s_1} p_{s_0 s_1}
 \label{couple3}
\end{gather}
\begin{gather}
 Q^*= \sum_{s_0=0}^{(z-1)} \sum_{s_1=0}^{(z-1)-s_0} \frac{(z-1)!}{s_0!s_1!((z-1)-s_1-s_0)!}
 (Q^*)^{s_0} \nonumber\\
  (P^*)^{s_1} (1-Q^*-P^*)^{(z-1)-s_0-s_1} q_{s_0 s_1}
 \label{couple4}
\end{gather}
The value $P_C$ and $Q_C$ at the root node $C$  as $n \rightarrow \infty$ is then given by
\begin{gather}
 P_C= \sum_{s_0=0}^{z} \sum_{s_1=0}^{z-s_0} \frac{z!}{s_0!s_1!(z-s_1-s_0)!} 
 (Q^{*})^{s_0}\nonumber\\
  (P^{*})^{s_1} (1-Q^{*}-P^{*})^{z-s_0-s_1} p_{s_0 s_1}
 \label{master5}
\end{gather}
\begin{gather}
 Q_C= \sum_{s_0=0}^{z} \sum_{s_1=0}^{z-s_0} \frac{z!}{s_0!s_1!(z-s_1-s_0)!} 
 (Q^{*})^{s_0} \nonumber\\
  (P^{*})^{s_1} (1-Q^{*}-P^{*})^{z-s_0-s_1} q_{s_0 s_1}
 \label{master6}
\end{gather}
The probabilities $P_C$ and $Q_C$ are the values deep inside the tree far away from the surface and are the estimate of the fraction of the $+1$ and $0$ spins respectively on a Bethe lattice. Magnetization per spin on a Bethe lattice is given by $m=2 P_C+Q_C-1$.

\subsection{Results}
\begin{figure}
\includegraphics[width=0.9\columnwidth]{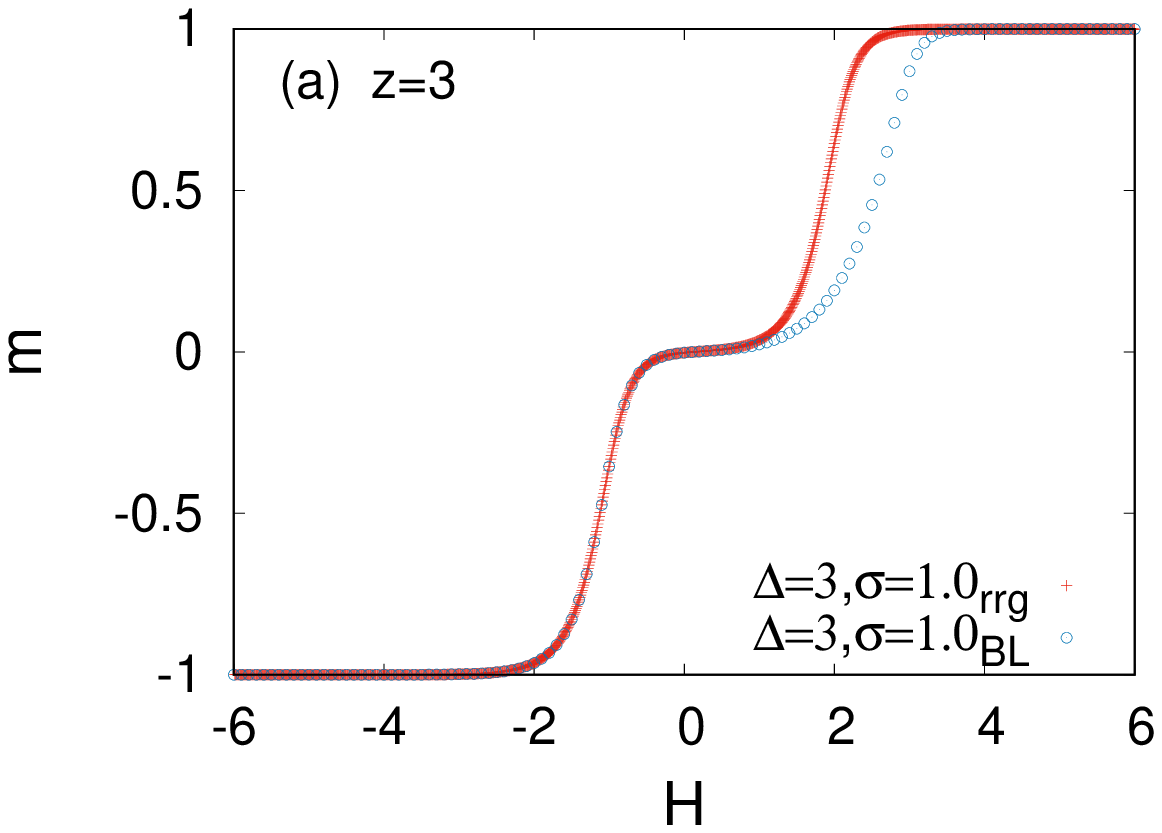}
    \includegraphics[width=0.9\columnwidth]{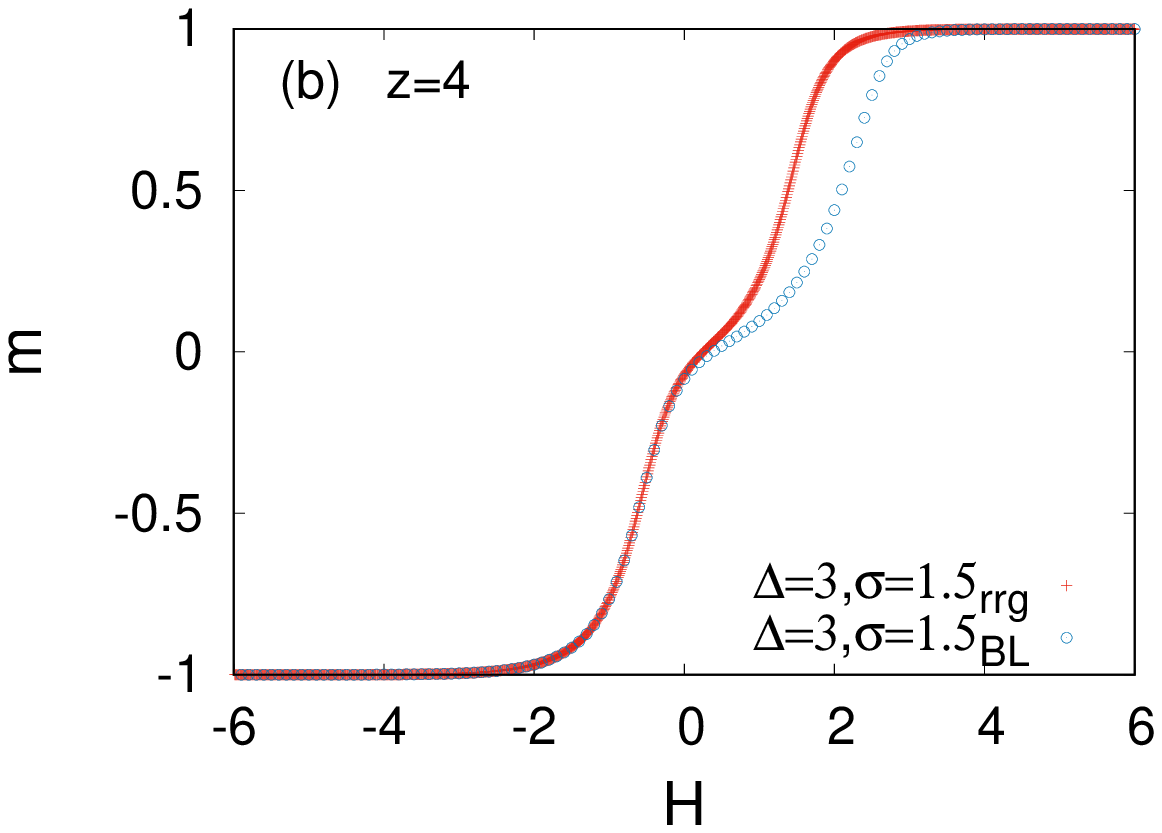}
   \caption{{Magnetization $m$ on a Bethe lattice (BL) (the blue (light gray) line of circles) as a function of external magnetic field $H$ for coordination numbers $z=3,4$ is compared with the value from simulations on a random regular graph (rrg) (the red (dark gray) line of plusses) for same value of $z,\Delta,\sigma$.}}
\label{mag}
\end{figure}

\begin{figure}
\includegraphics[width=0.9\columnwidth]{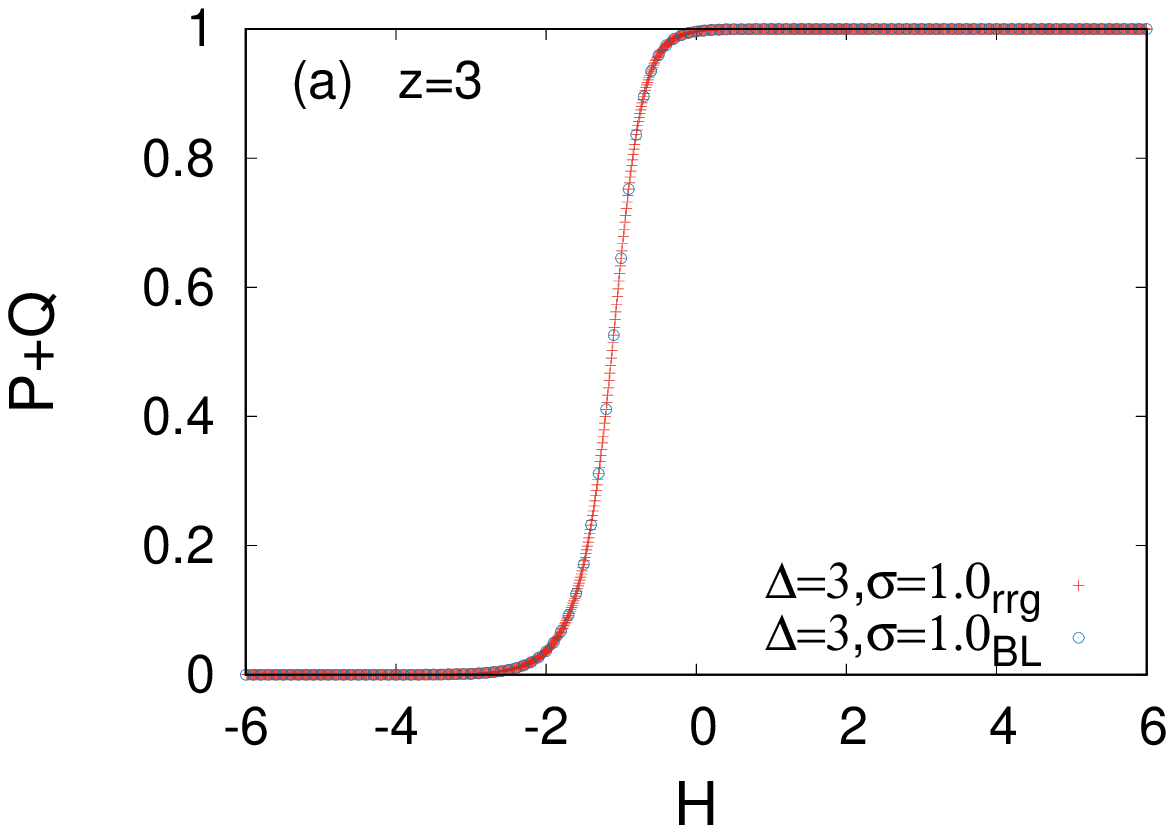}
    \includegraphics[width=0.9\columnwidth]{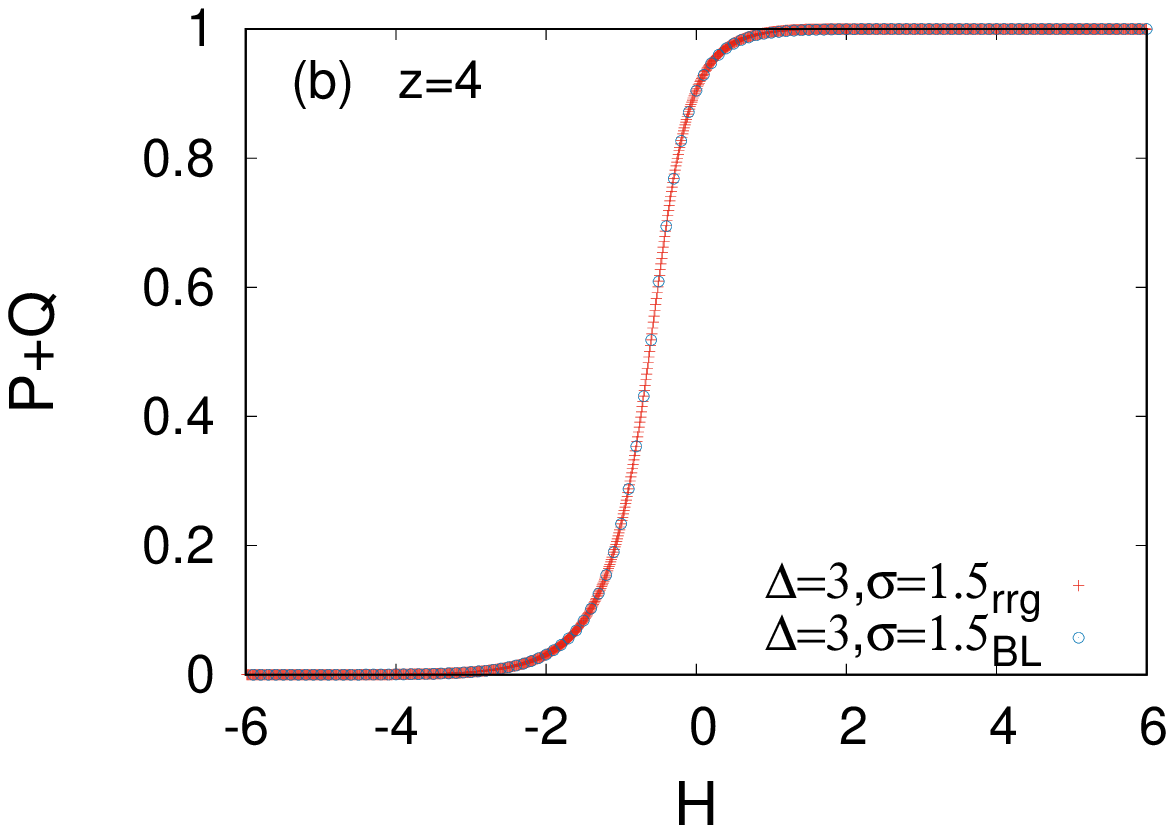}
   \caption{{$P_C + Q_C$ as a function of external magnetic field $H$ from the Bethe lattice (BL) (the blue (light gray) line of circles) calculations is compared with the sum of the  probabilities of $+1$ spin state and $0$ spin state from the simulations on random regular graphs (rrg) (the red (dark gray) line of plusses). The two plots overlap showing a good match between the analytical calculation and simulations.}}
\label{fraction}
\end{figure}

The solution obtained above matches with the simulations on the random regular graph of the same coordination number for negative values of the external field. In Fig. \ref{mag} we plot the magnetization $m$ obtained via the above calculation with the simulation results on a random regular graph for $z=3$, $\Delta=3$, $\sigma=1$ and for $z=4$, $\Delta=3$, $\sigma=1.5$. The behaviour is similar for other sets of ($z$, $\Delta$, $\sigma$) as well (see Appendix \ref{appC}). We also plot the sum of $P_C$ and $Q_C$ along with their values from simulations on a random regular graph of the same coordination number in Fig. \ref{fraction}.
  
\begin{figure}
\includegraphics[width=0.45\columnwidth]{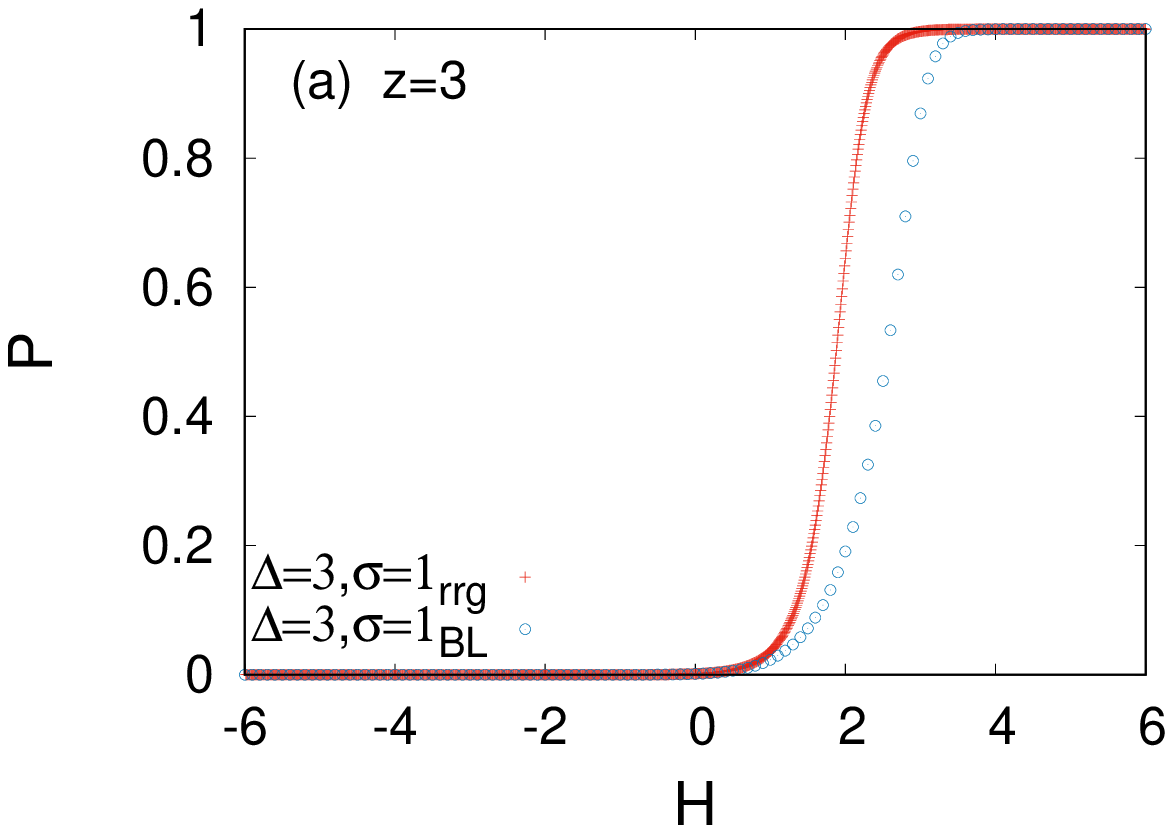}
    \includegraphics[width=0.45\columnwidth]{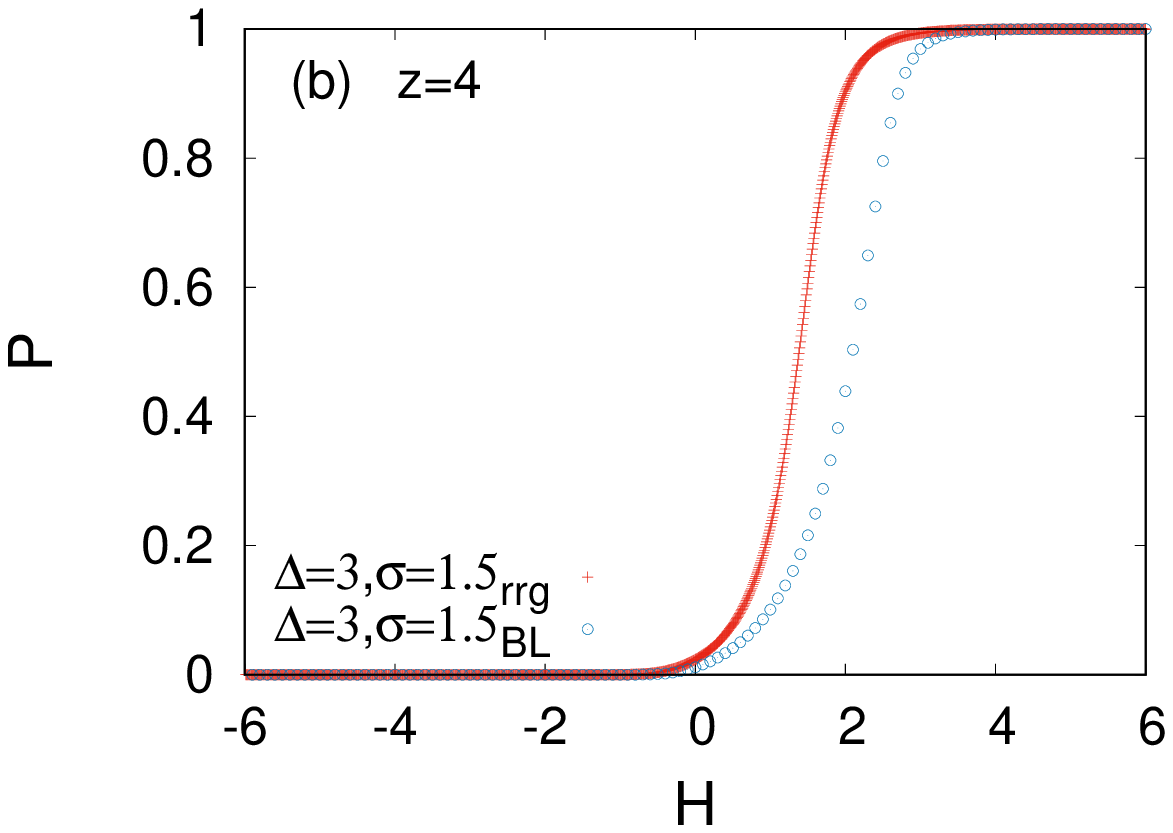}
   \includegraphics[width=0.45\columnwidth]{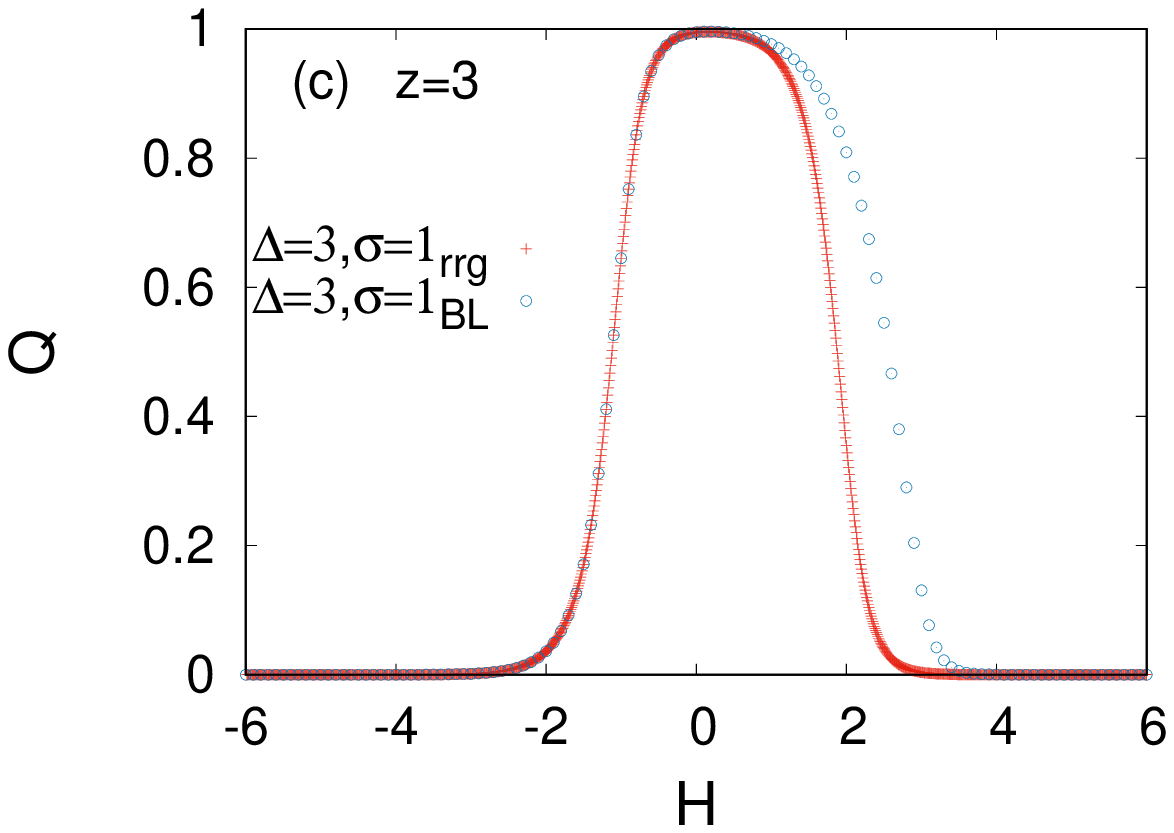}
    \includegraphics[width=0.45\columnwidth]{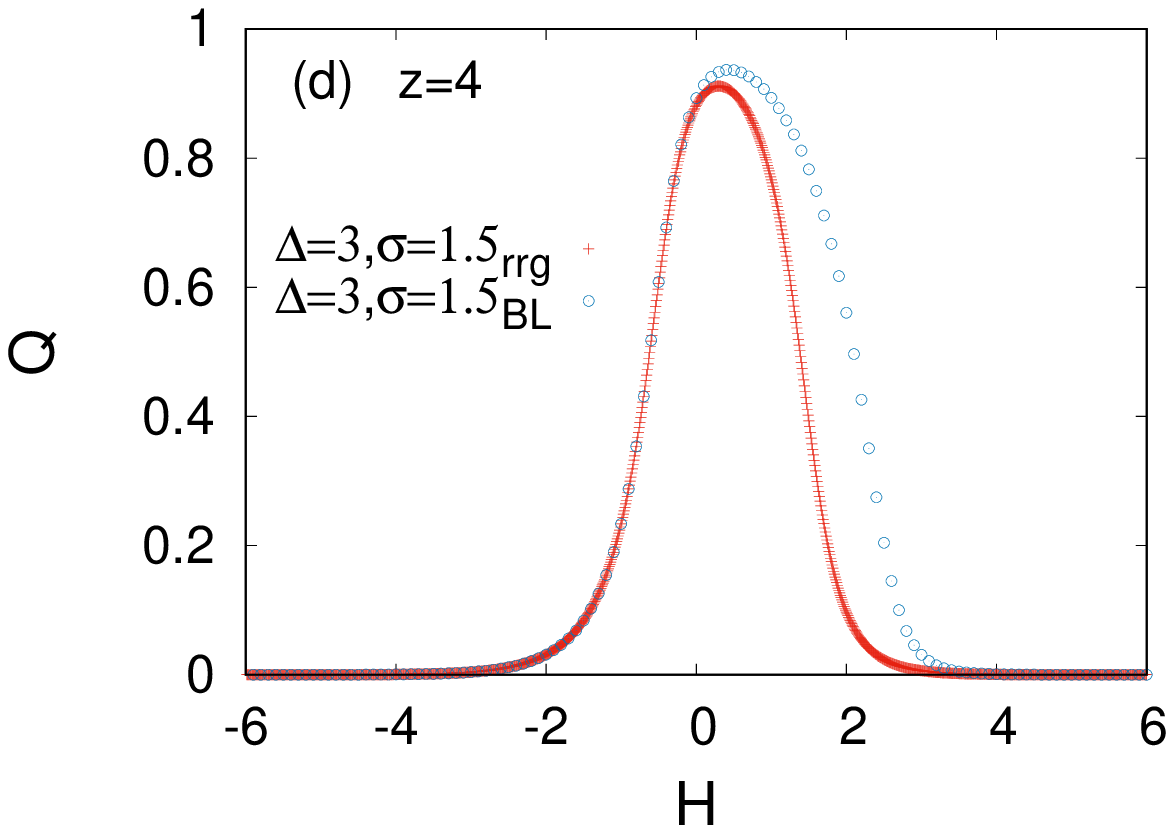}
\caption{$P_C$ and $Q_C$ as a function of external magnetic field $H$ for $z=3$ and $z=4$ is shown. $Q_C$ is overestimated and $P_C$ is underestimated for positive $H$ in the Bethe lattice (BL) (the blue (light gray) line of circles calculations when compared with the value on a random regular graph (rrg) (the red (dark gray) line of plusses).}
\label{com}
\end{figure}
As shown in Fig. {\ref{mag}},  $m$ matches with the simulated 
data for $H<0$ but deviates for positive $H$. 
For spin-$1$, we find an underestimation for $P_C$ and an overestimation for $Q_C$, as $H$ increases. In principle once a spin at level $r$ flips, it can change the configuration of the spins at level $r+1$, which has been ignored in the calculations on the Bethe lattice. This though does not change the value of the probabilities at node $C$ for RFIM as a change in $r+1$ cannot further change the spin at level $r$ (it is already in $+1$ state). But the same is not true for RFBCM model. For example, let us assume that a spin at level $r$ attains a $0$ value due to its descendants and in turn makes its descendants flippable and hence they  change their state. It is now  possible that its descendant spins in level $r+1$ produce enough interaction field at the spin that now the same spin at level $r$ might prefer to be in $+1$ state. 
Ignoring this results in the underestimation of $P_C$ and overestimation of $Q_C$ as $H$ is increased (see Fig. \ref{com}). The sum $P_C+Q_C$ is correctly calculated in the Bethe lattice calculation. In Fig. \ref{fraction} the plot of $P_C+Q_C$ on a Bethe lattice and the one obtained in simulations on a random regular graph overlap for all values of $H$. 

A solution on the Bethe lattice under the abelian assumption fairly accurately calculates the lower hysteresis loop. Since the $m$ vs $H$ curves for random regular graphs as shown in Fig. \ref{hystsim} for positive and negative $H$ can be superimposed on each other by reversing the sign, the solution obtained in this section gives the solution of the problem on a Bethe lattice. On the Bethe lattice for $ z \ge 3$, $P^*$ and $Q^*$ become multivalued when $H$ is near the transition for $\sigma< \sigma_c$. The hysteresis is thus discontinuous for $\sigma<\sigma_c$ and continuous for $\sigma >\sigma_c$ $\forall z \ge 3$ also on a Bethe lattice. The value of $\sigma_c$ depends on $z$ and $\Delta$. The value of $\sigma_c$ calculated from the Bethe lattice calculations is found to be consistent with the estimate from the simulations on a random regular graph. We tabulate the value of $\sigma_c$ from simulations and the Bethe lattice calculations for a few values of $\Delta$ for $z=4$ in Table.1.

\begin{table}[H]
\begin{center}
\begin{tabular}{|c|c|c|c|}
\hline
$z$ & $\Delta$ & $\sigma_c(simulations)$ & $\sigma_c(Bethe\ lattice)$  \\
\hline
4 & -1 & 1.8(2) & 1.9(1) \\\hline
4 & 1  & 1.8(2)  & 1.9(1) \\\hline
4 & 3 & 0.9(2) & 1.1(1) \\\hline
4 &  5 & 0.9(2) & 1.1(1) \\\hline
\end{tabular}
\end{center}
\caption{Critical value of $\sigma$ at which the hysteresis plot become continuous, as estimated in the simulations and calculations on the Bethe lattice.}
\label{tab:1}
\end{table}

\section{Discussions}
\label{sc}
The zero temperature RFIM has been studied extensively as it has been successful in explaining the mechanism of hysteresis for many experimental systems \cite{noise}. In this paper, we studied the zero temperature spin-$1$ RFBCM numerically on a random regular graph and extended the earlier calculation of the magnetization for RFIM  on a Bethe lattice to RFBCM. We showed that the  Bethe lattice calculations match with the simulation results on a random regular graph for negative $H$ in spite of the absence of the abelian property of the Glauber dynamics for RFBCM. 

It would be interesting to expand this study to other higher spin ferromagnetic random field models. In particular zero temperature spin-$1$ random field Blume Emery Griffiths model (RFBEGM), which differs from RFBCM by the presence of an extra bi-quadratic interaction \cite{beg}. The zero temperature RFBEGM  was numerically studied on a square lattice in \cite{ortin2} as a model for martensite transitions by considering the $0$ and $\pm 1$ spin states as austenite and two martensite states respectively. They observed that the model exhibits double hysteresis without RPM \cite{ortin2}. In experiments, RPM is almost always observed in thermoelastic martensite transitions \cite{ortin1}. In this paper, we show that a spin-$1$ random field model can also exhibit RPM. This suggests that the presence of the biquadratic term results in the loss of RPM property for a spin-$1$ ferromagnet. In general, the conditions under which a system exhibit RPM are not yet fully understood \cite{abhishek,libal}. Critical properties of the zero temperature non equilibrium steady state of spin-$1$ models like RFBCM and RFBEGM is largely unexplored and it would be interesting to study these models in different dimensions. 

\section{Acknowledgements}
We thank Deepak Dhar for discussions. A.K thanks NISER, Bhubaneswar for local hospitality when the part of this work was done.

\appendix 

\section{}
\label{appA}
\subsection{Thermalization time}
	
\begin{figure*}[h!]
    \includegraphics[width=0.65\columnwidth]{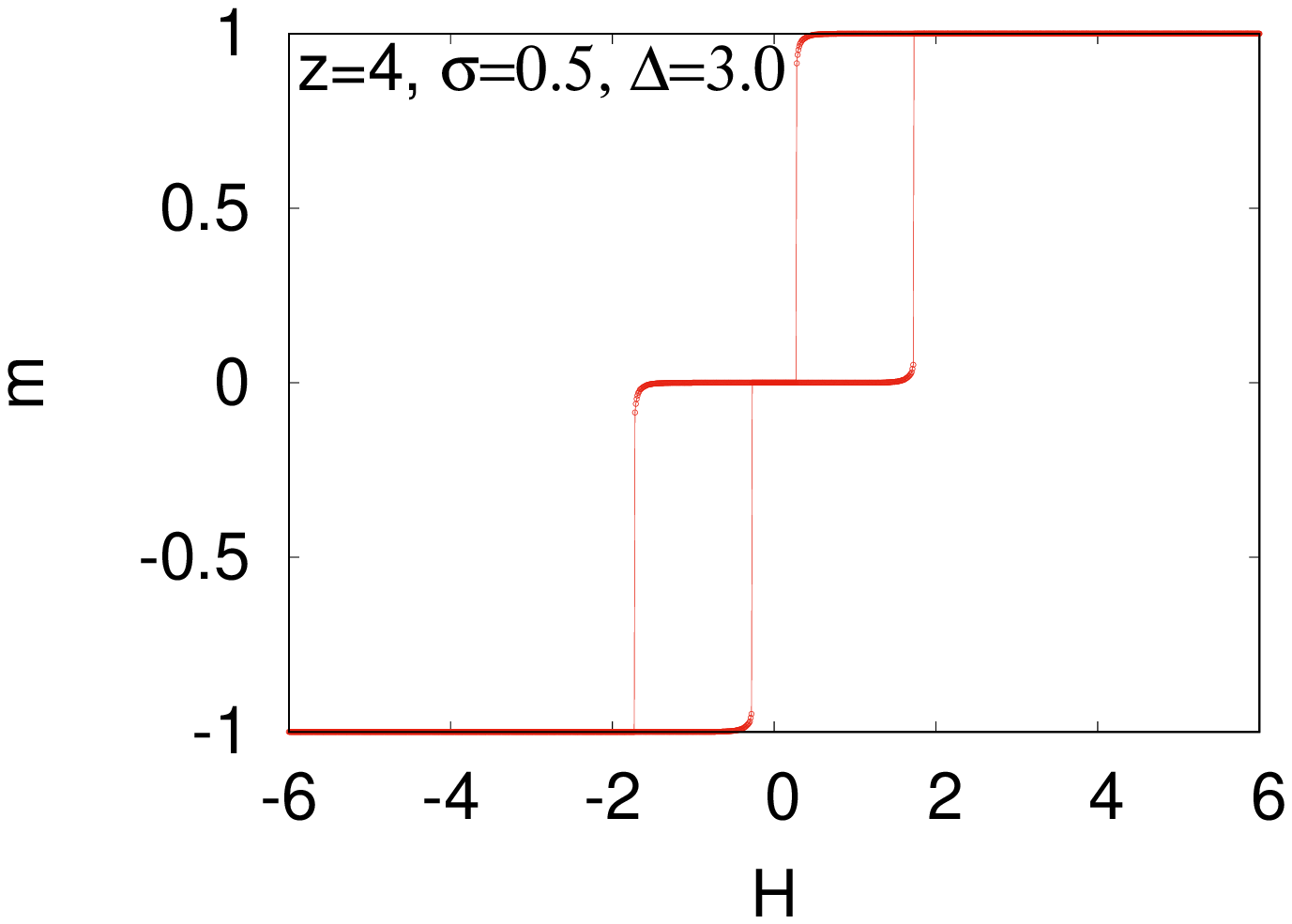}
    \includegraphics[width=0.65\columnwidth]{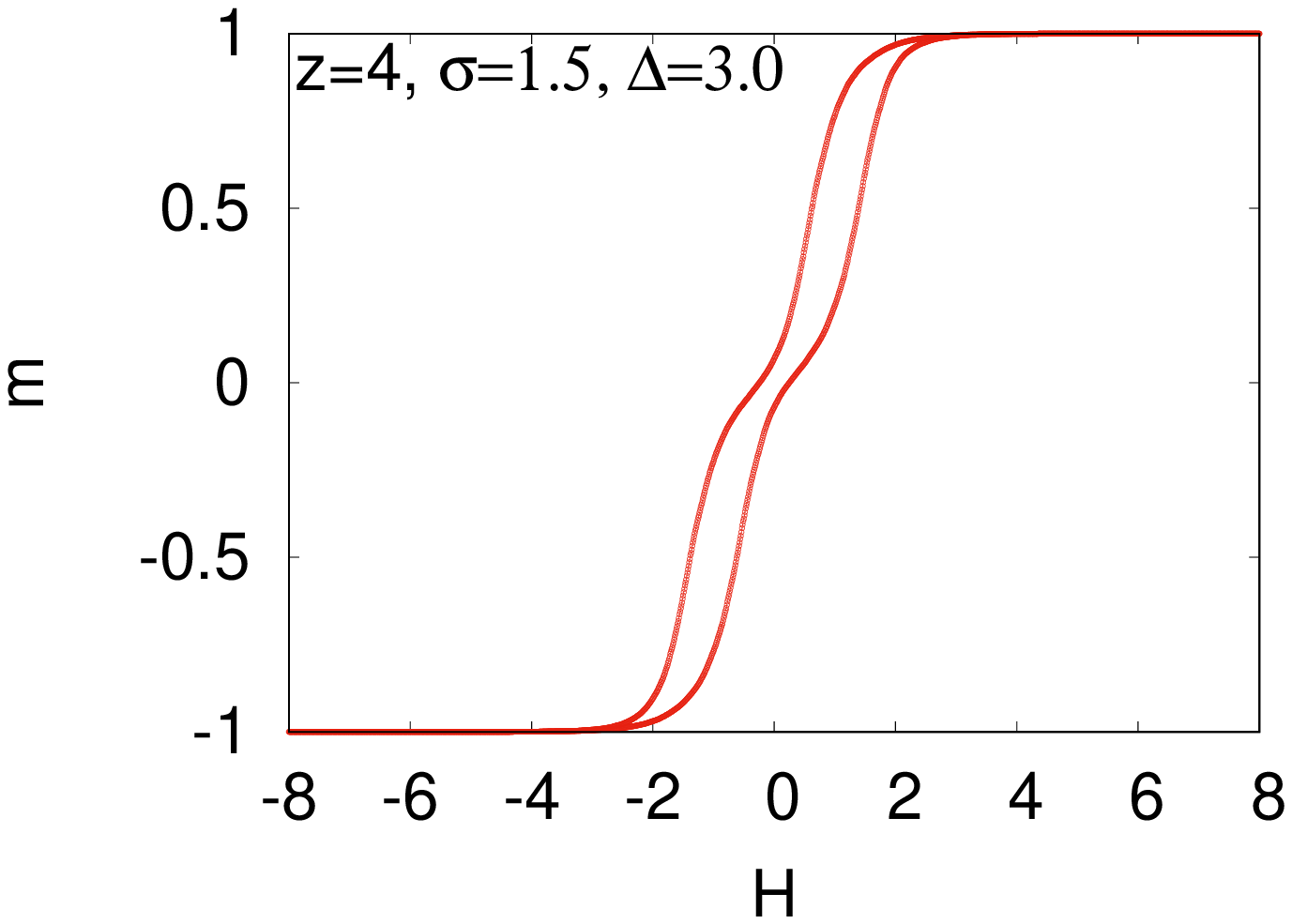}
    \includegraphics[width=0.65\columnwidth]{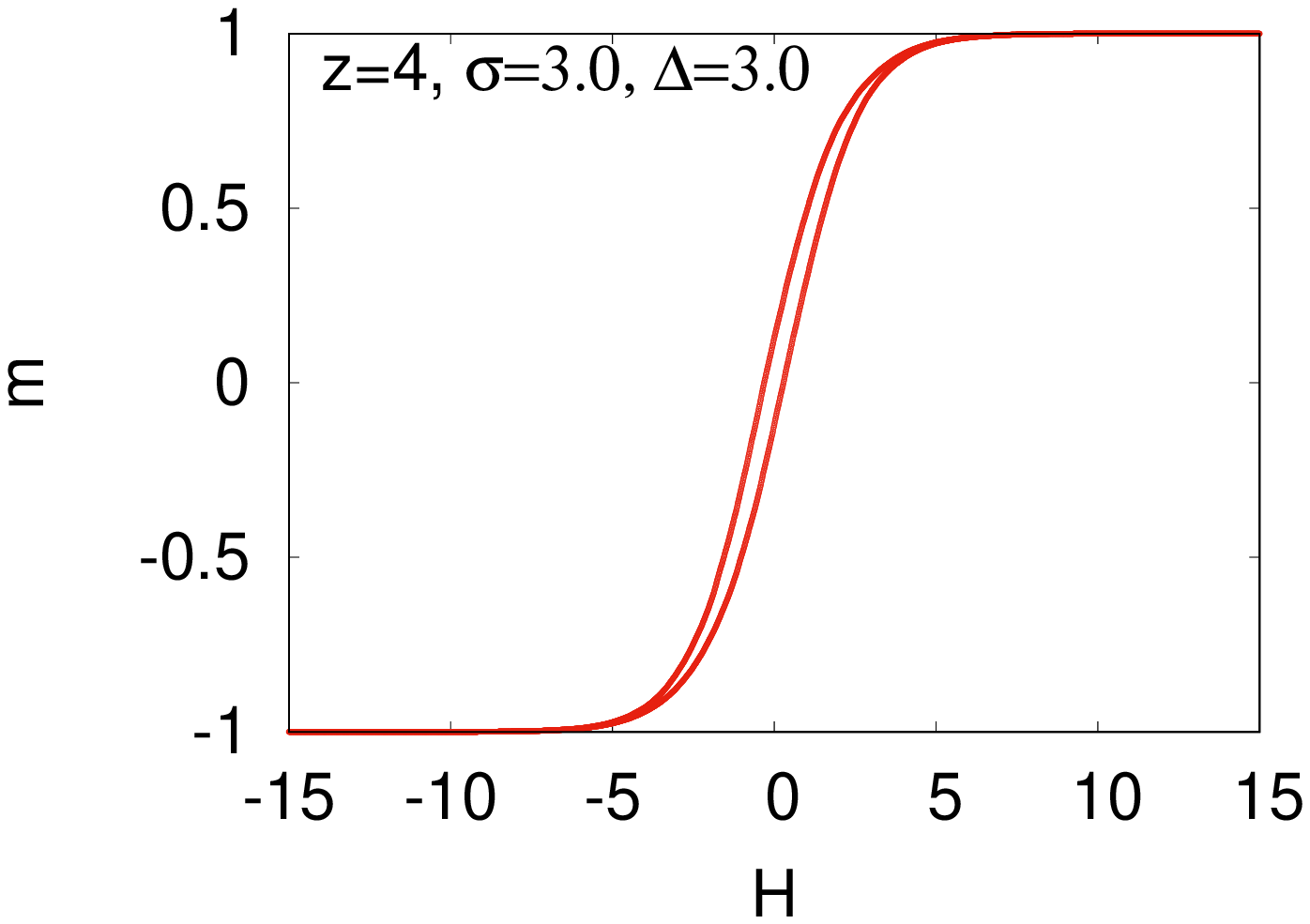}
    \includegraphics[width=0.65\columnwidth]{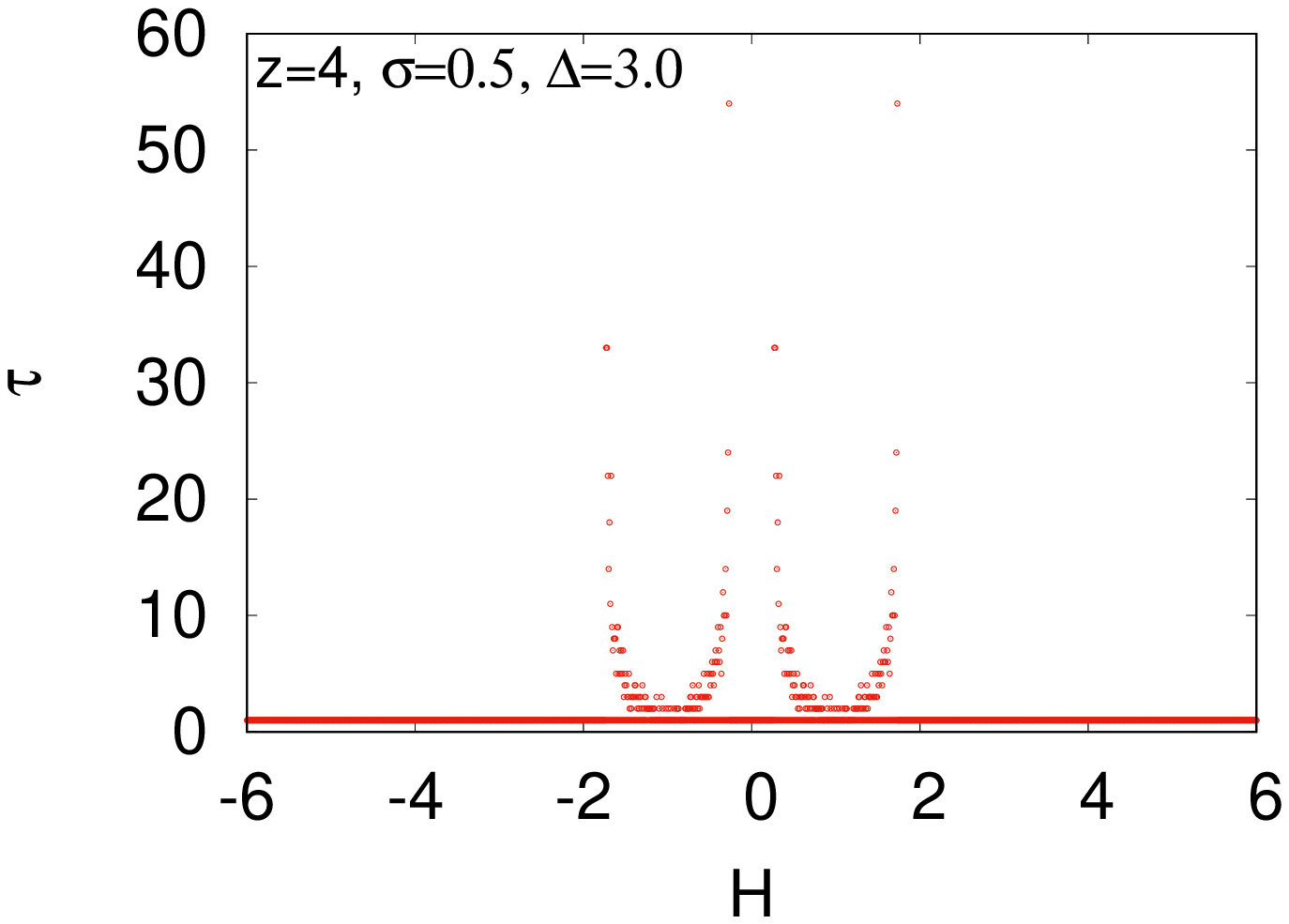}
    \includegraphics[width=0.65\columnwidth]{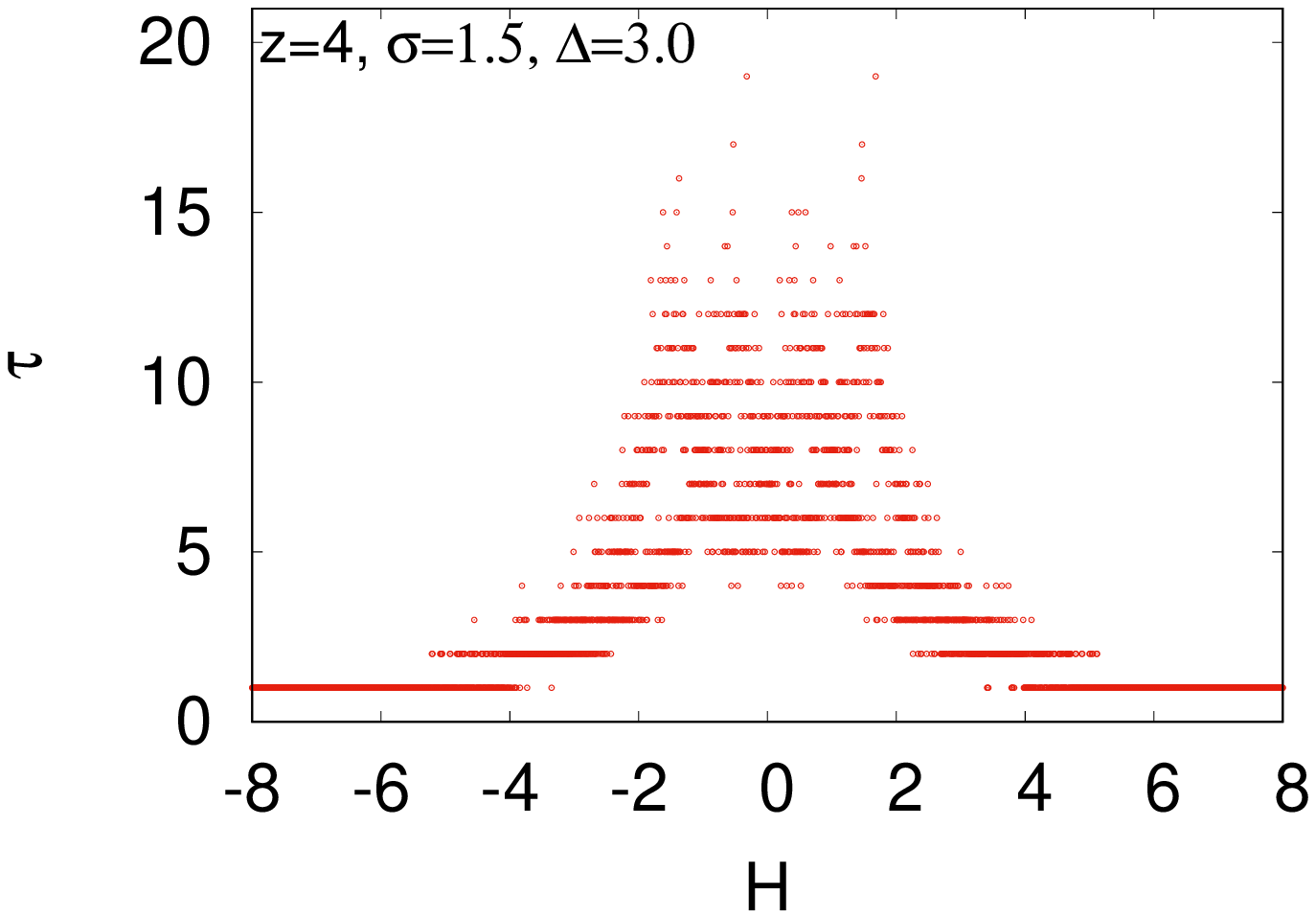}
    \includegraphics[width=0.65\columnwidth]{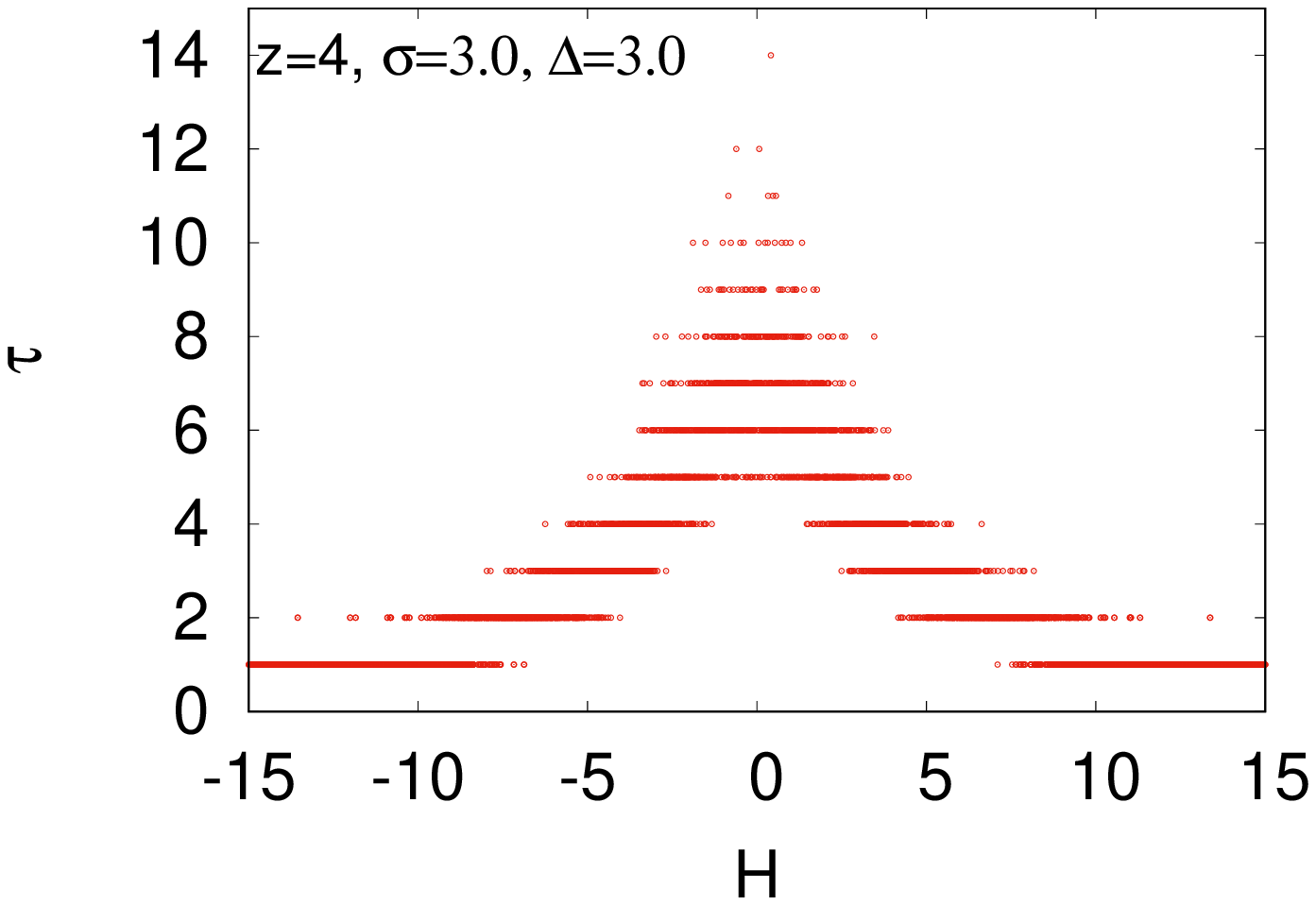}
\caption{ (Appendix A.1) Thermalization time ($\tau$) plots are shown against magnetization ($m$) plots for z = 4, $\Delta = 3$ for different values of $\sigma = 0.5, 1.5, 3$. }
\label{A1}
\end{figure*}

	Thermalization time $\tau$ is defined as the number of Monte Carlo steps taken by the system to reach a steady state, for a fixed value of the parameters like the external field. One sweep of the entire system is taken as the one Monte Carlo step. Hence $\tau$ for a given $H$  is the total number of times we run through the whole lattice looking for possible spin flips, until we reach a spin configuration that yields no spin flips at any of the lattice sites. We plot $\tau$ as a function of the external field $H$ against the magnetization plots for $z = 4$, $\Delta = 3$ and for different values of $\sigma = 0.5, 1.5, 3$ in FIG. \ref{A1}. The $\tau$ as expected increases near the transition.

\begin{figure*}[h!]
\includegraphics[width=\columnwidth]{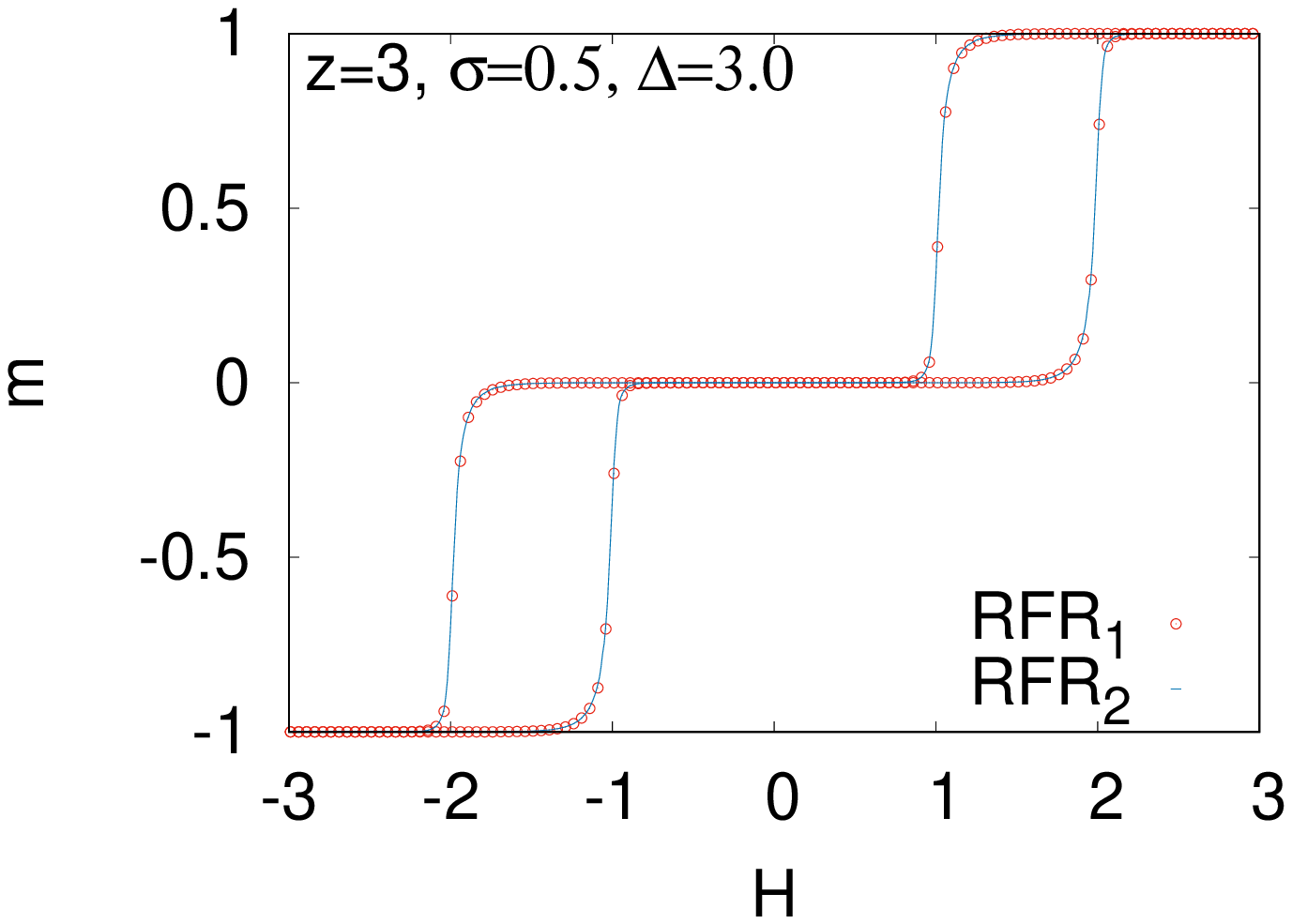}
    \includegraphics[width=\columnwidth]{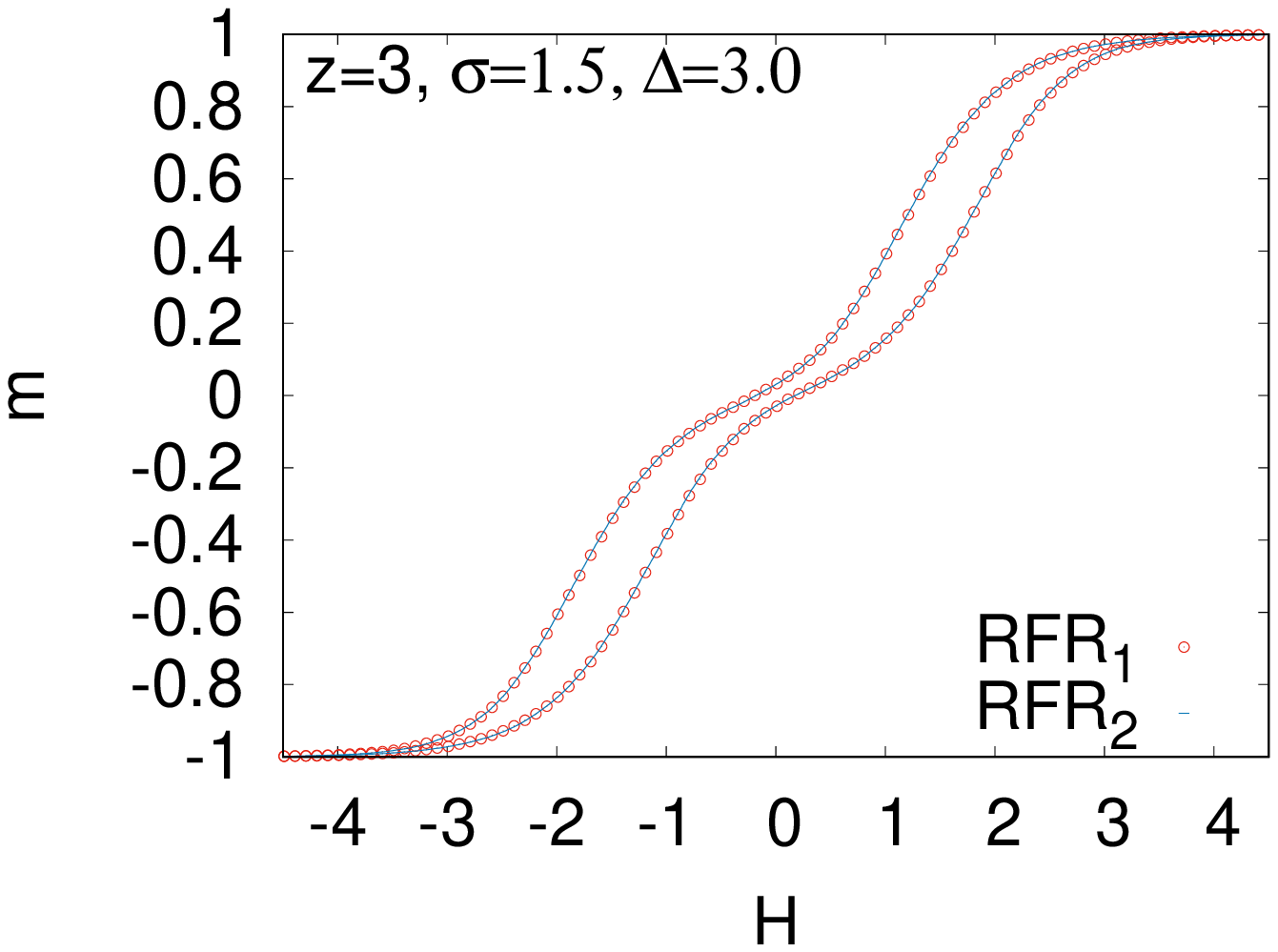}
\caption{(Appendix A.2) The magnetization plots of RFBCM obtained for two different random field realizations (RFR$_1$ and RFR$_2$) are indistinguishable at the scale of the graph, as shown for two different values of $\sigma = 0.5, 1.5$.}
\label{A3}
\end{figure*}
\subsection{Comparing magnetization plots of different random field realizations}
To show that for large system sizes the change in magnetization 
from one realization of disorder to the other is not noticeable,  we plot the magnetization for $z = 4$, $\Delta = 3$, for two different random field realizations for $\sigma = 0.5$ and $1.5$. For both of the values of $\sigma$, the plots obtained from different random field realizations (RFR$_1$ and RFR$_2$) are indistinguishable at the scale of the graph (as shown in FIG.\ref{A3}).
\subsection{Magnetization plots of different increments in the field}
\begin{figure}
\includegraphics[width=0.9\columnwidth]{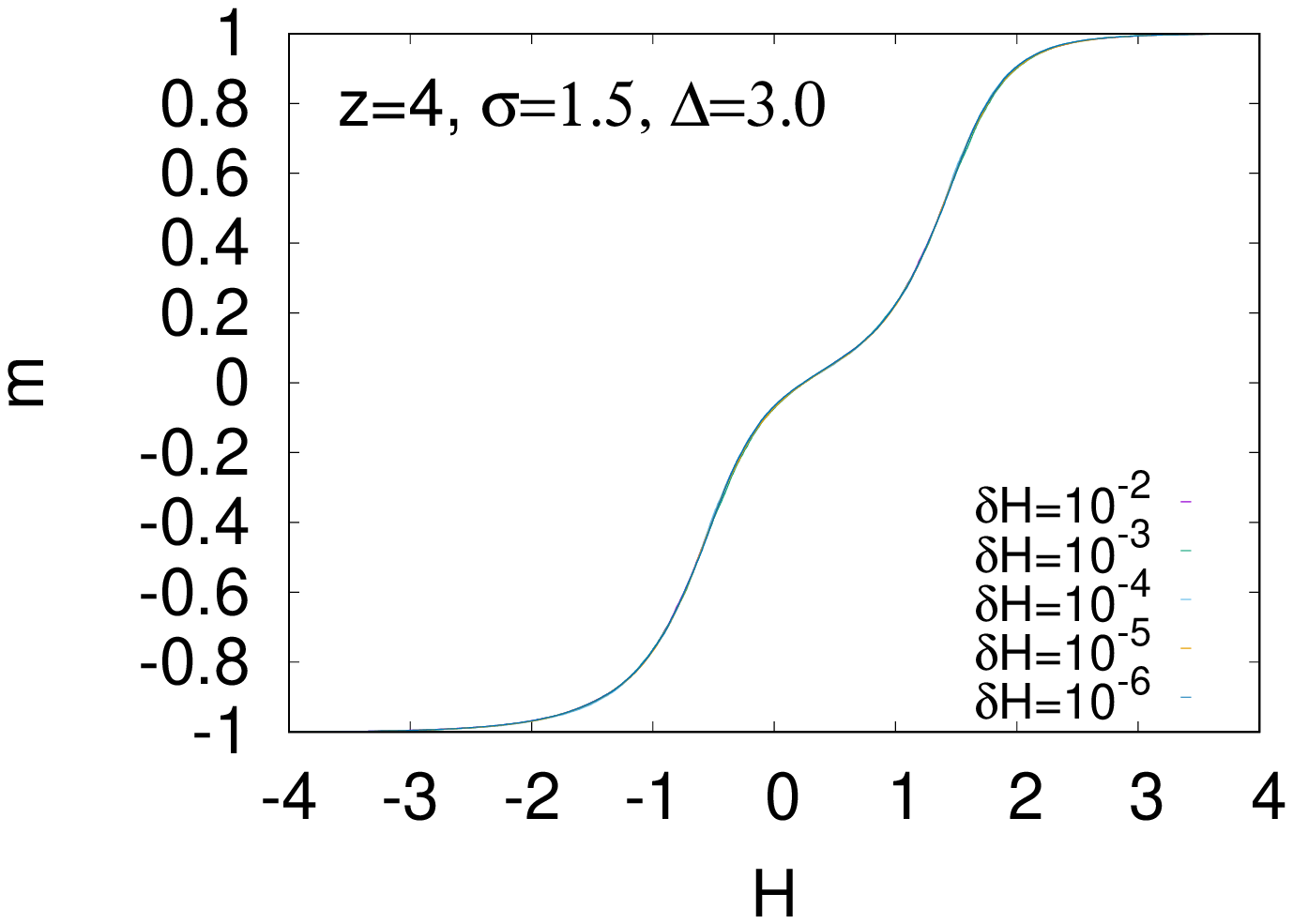}
   \caption{(Appendix A.3) The magnetization plots of RFBCM obtained for different values of the increments in the external field ($10^{-2}$, $10^{-3}$, $10^{-4}$, $10^{-5}$, $10^{-6}$) that are indistinguishable at the scale of the graph.}
\label{A2}
\end{figure}

The magnetization for $z = 4$, $\Delta = 3$ and $\sigma = 1.5$ obtained for different values of the increments in the external field ($\delta H$ =$10^{-2}$, $10^{-3}$, $10^{-4}$, $10^{-5}$, $10^{-6}$) are plotted in Fig. \ref{A2}. All the plots are indistinguishable. Hence we have taken $\delta H =10^{-2}$ in our simulations.
\section{Hysteresis plots for various values of $z$,$\Delta$ and $\sigma$}
\label{appB}
\begin{figure*}[h!]
\includegraphics[width=0.65\columnwidth]{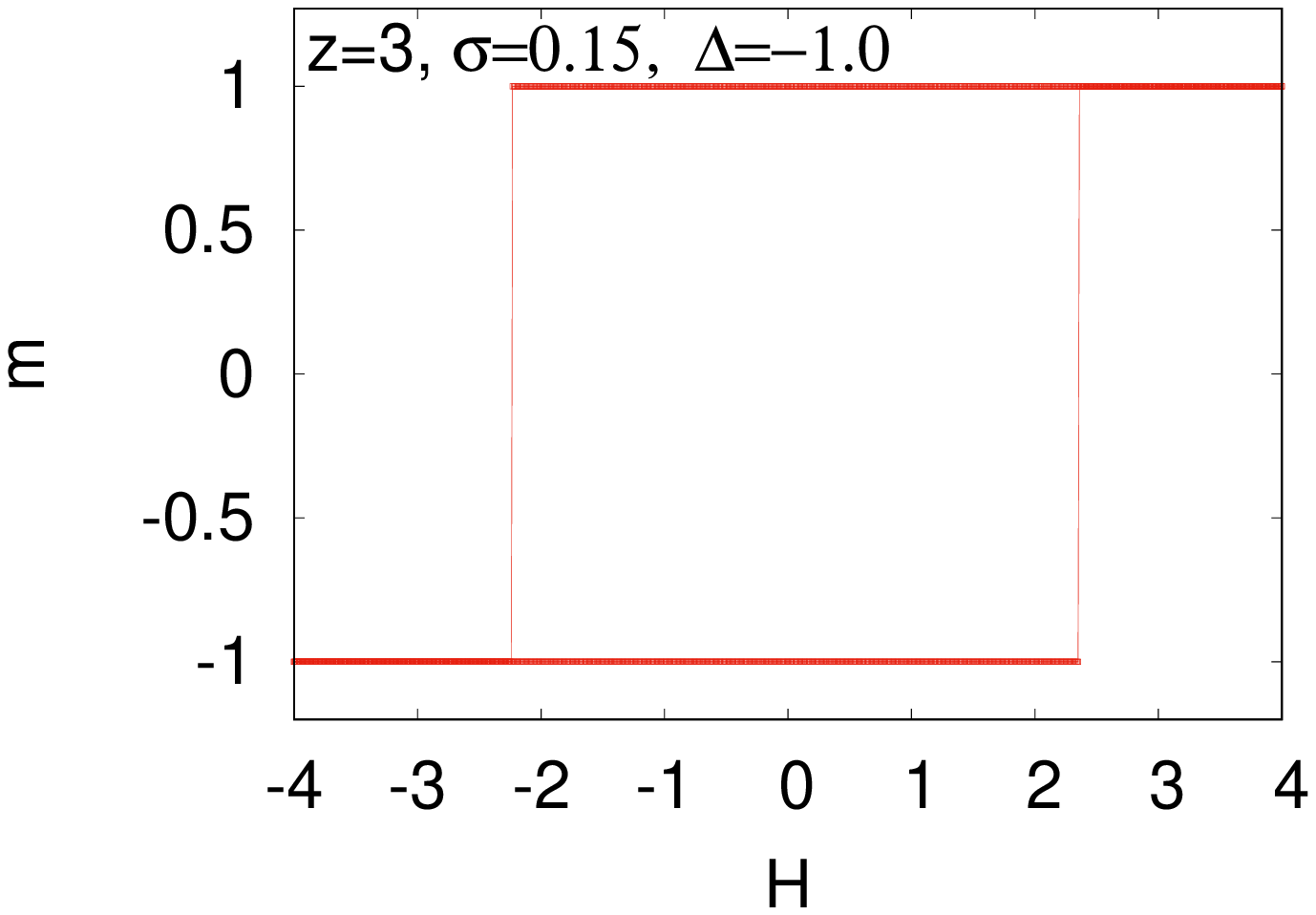}
\includegraphics[width=0.65\columnwidth]{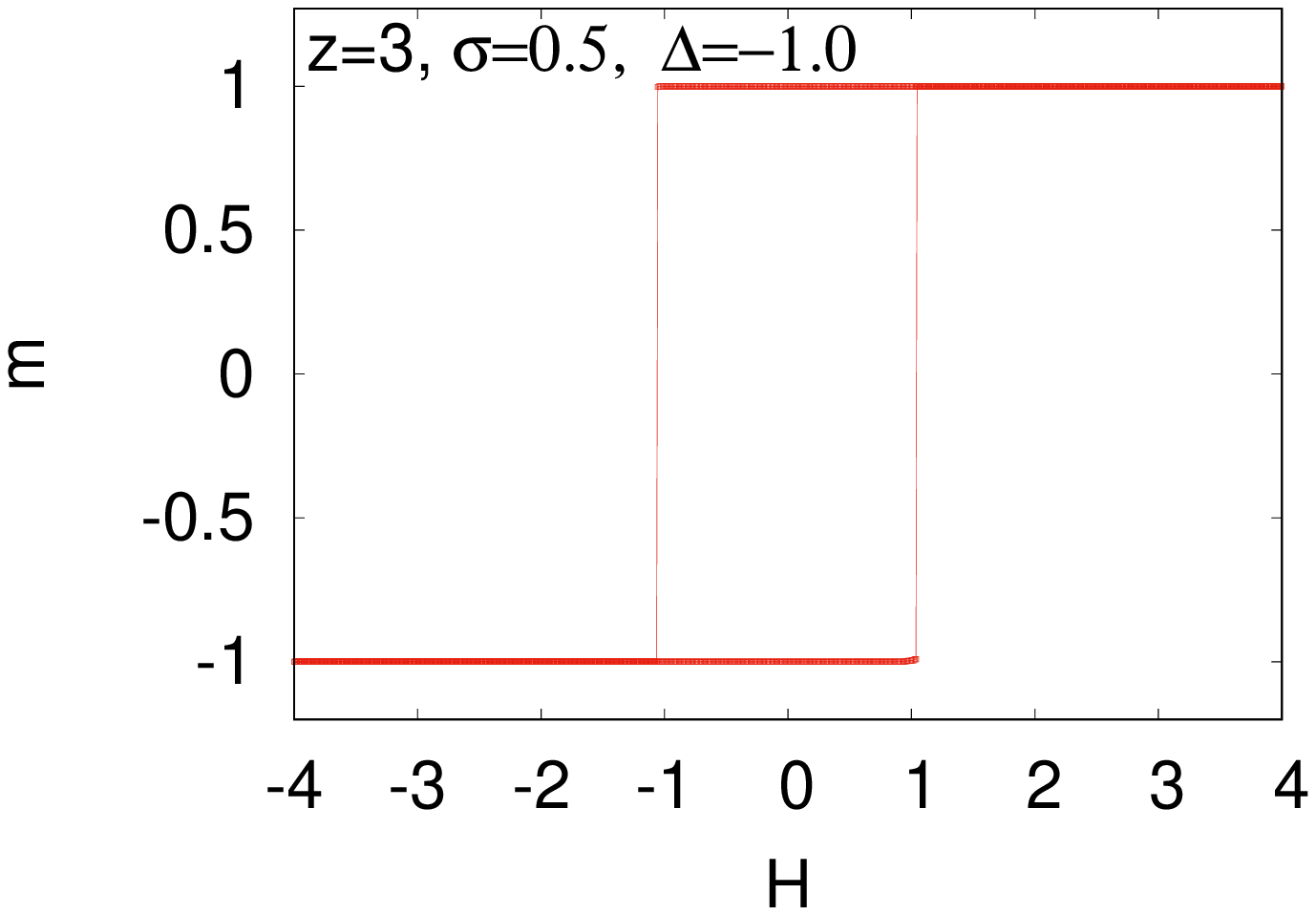}
\includegraphics[width=0.65\columnwidth]{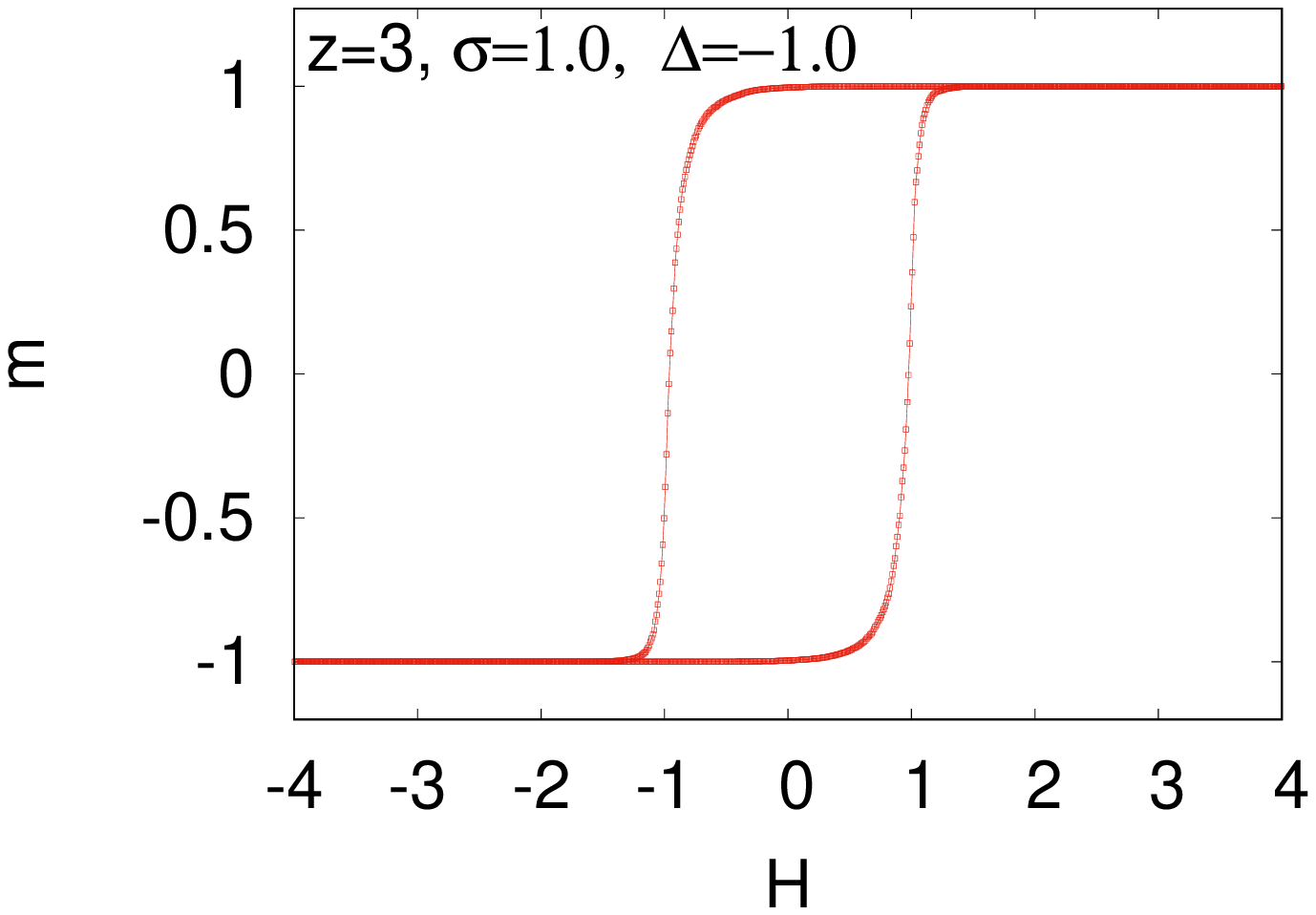}
\includegraphics[width=0.65\columnwidth]{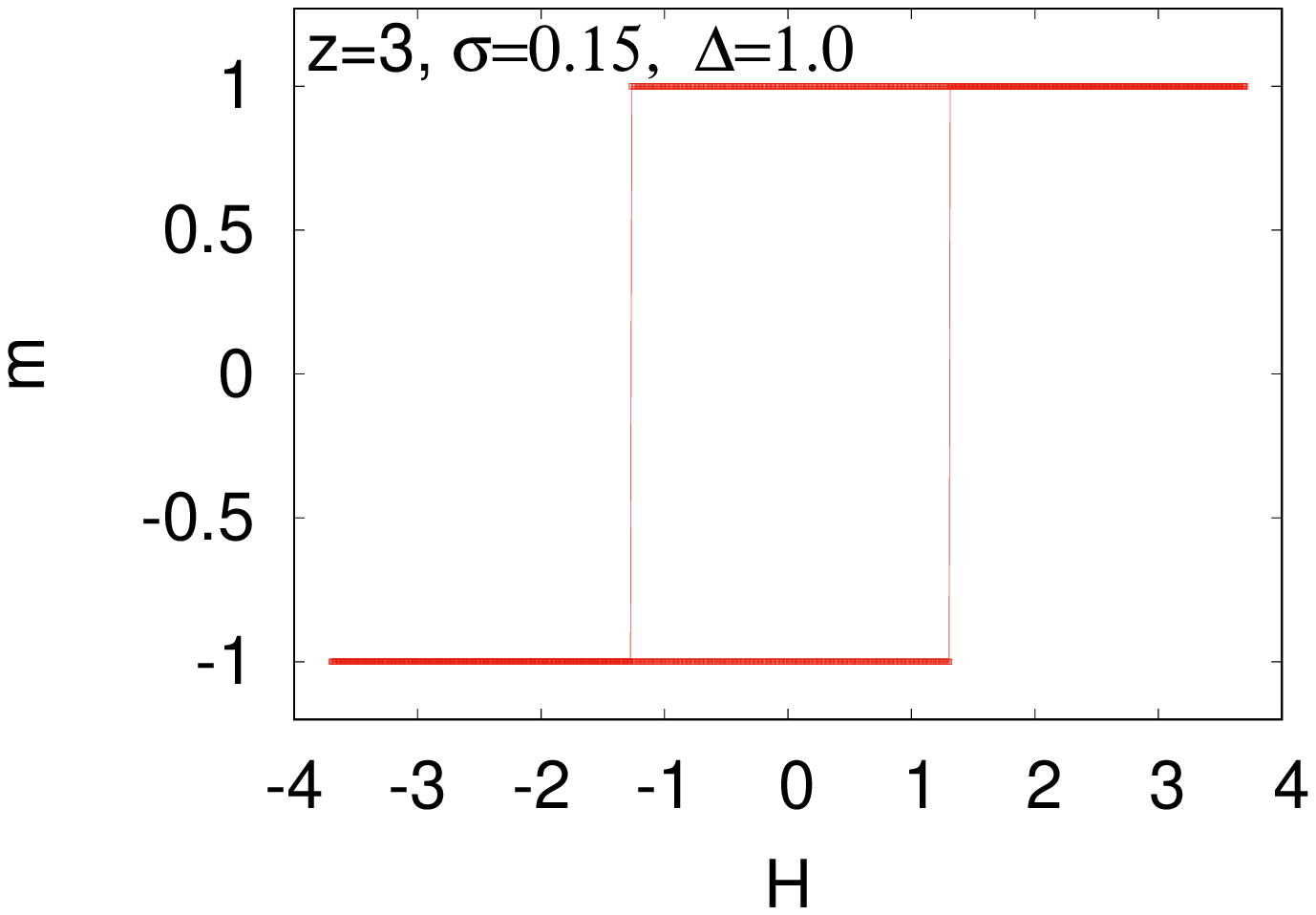}
\includegraphics[width=0.65\columnwidth]{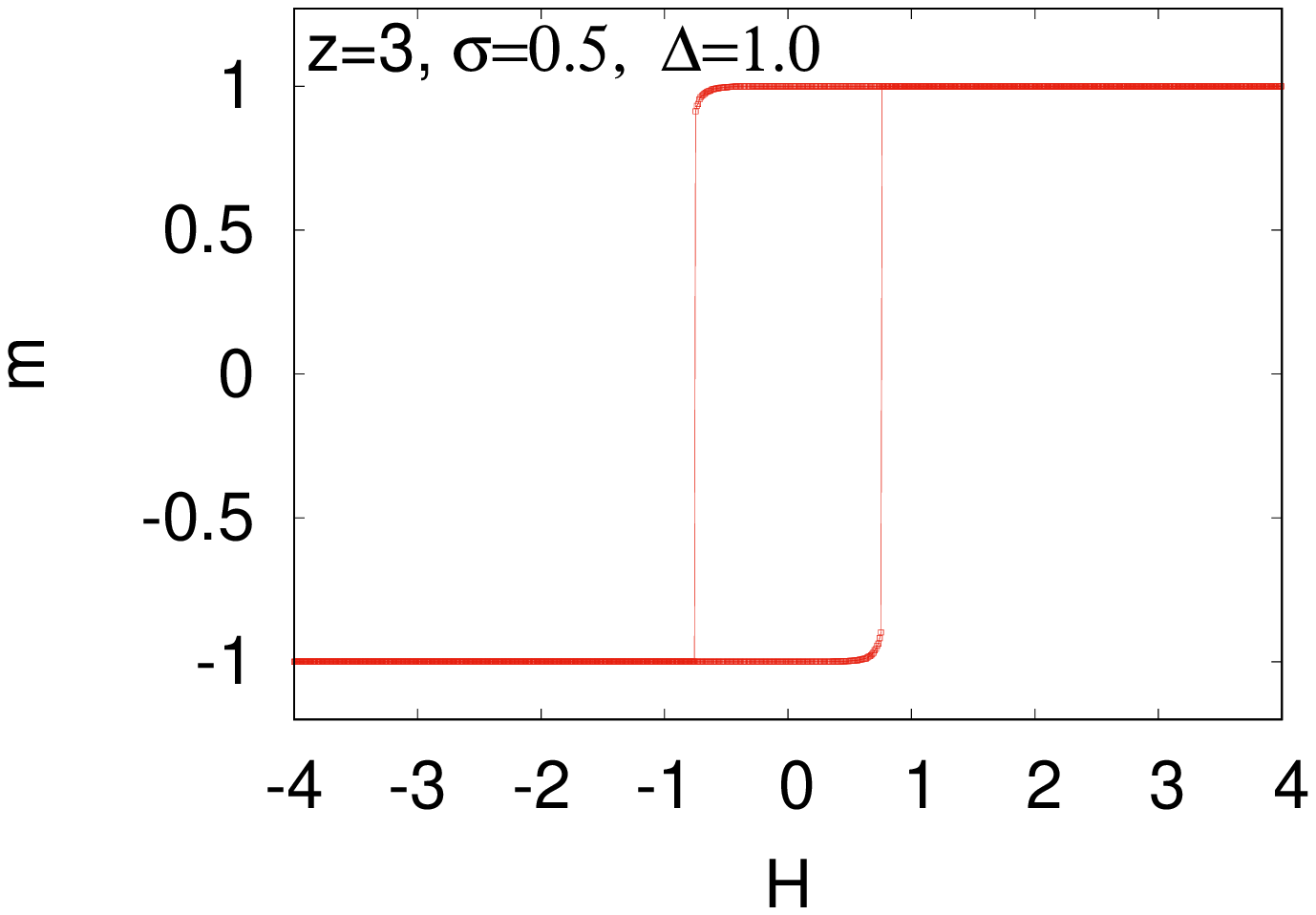}
\includegraphics[width=0.65\columnwidth]{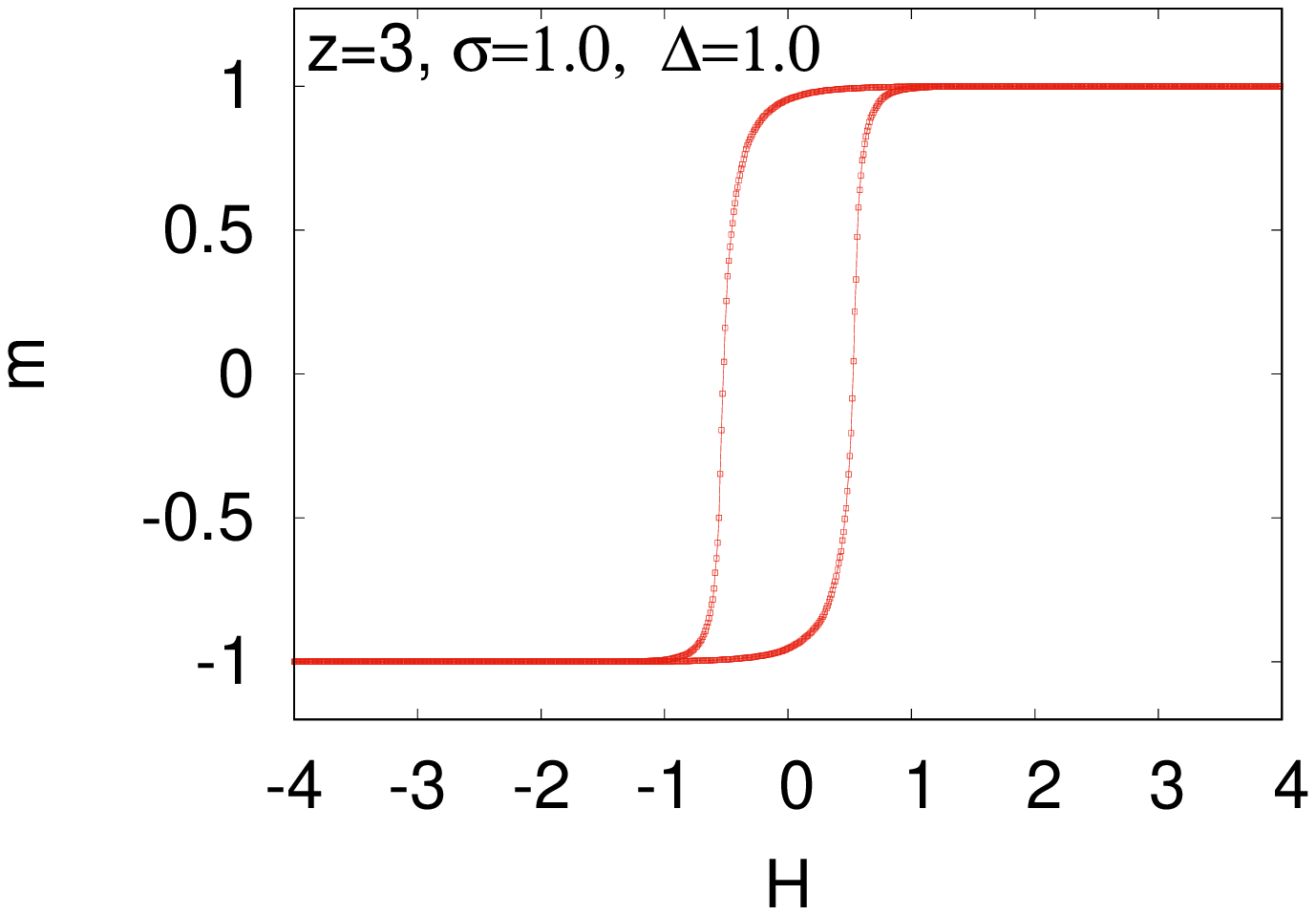}
\includegraphics[width=0.65\columnwidth]{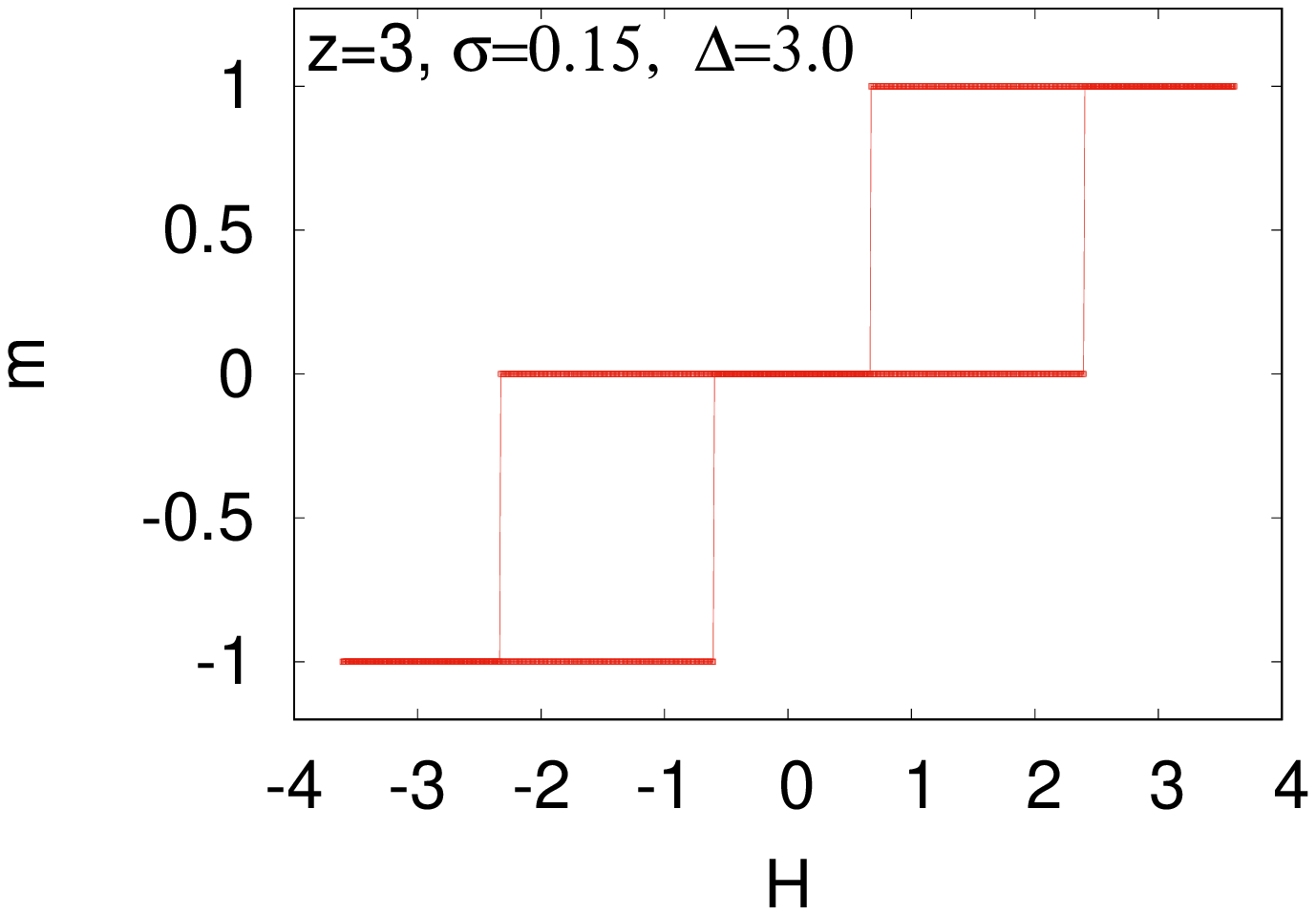}
\includegraphics[width=0.65\columnwidth]{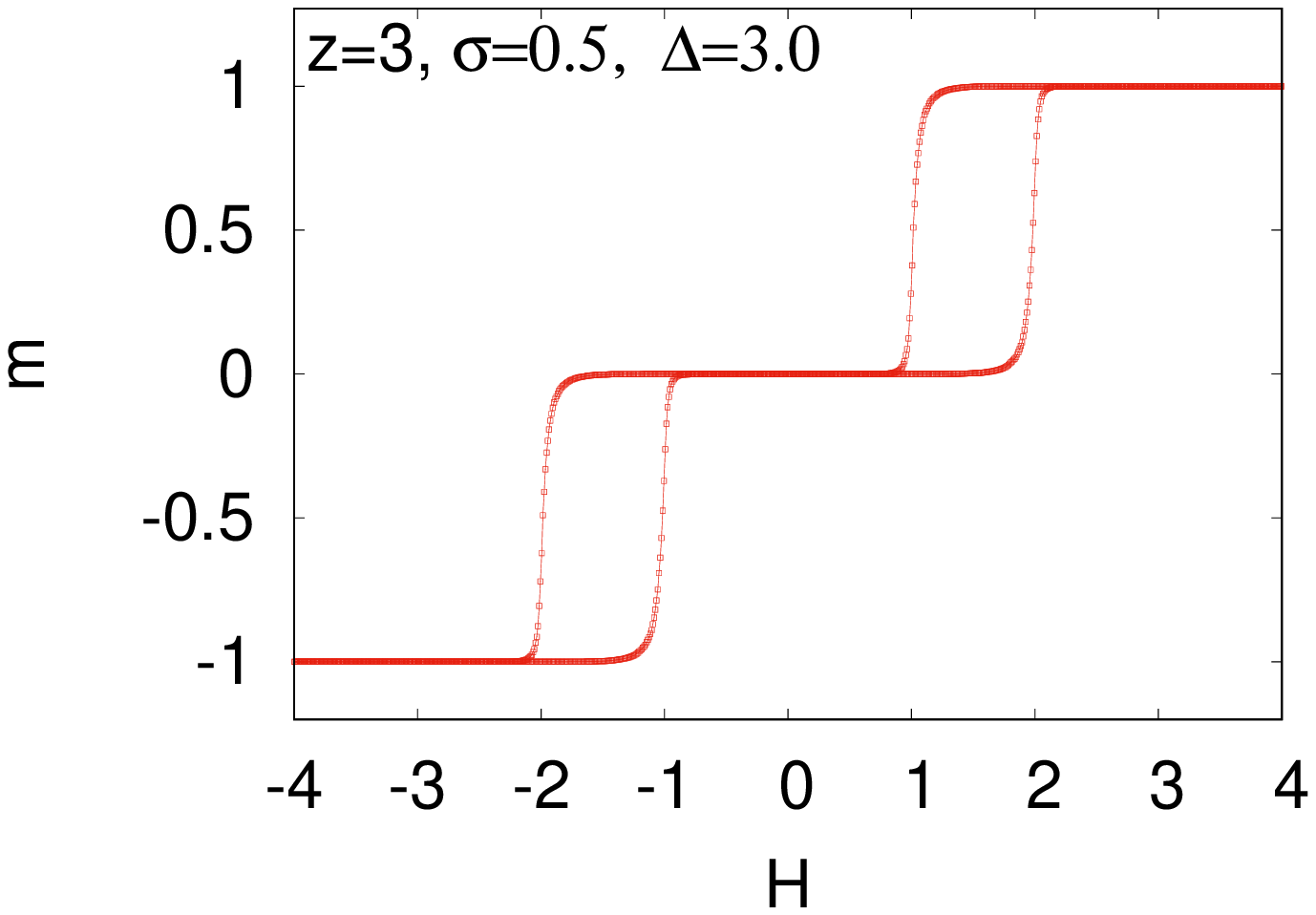}
\includegraphics[width=0.65\columnwidth]{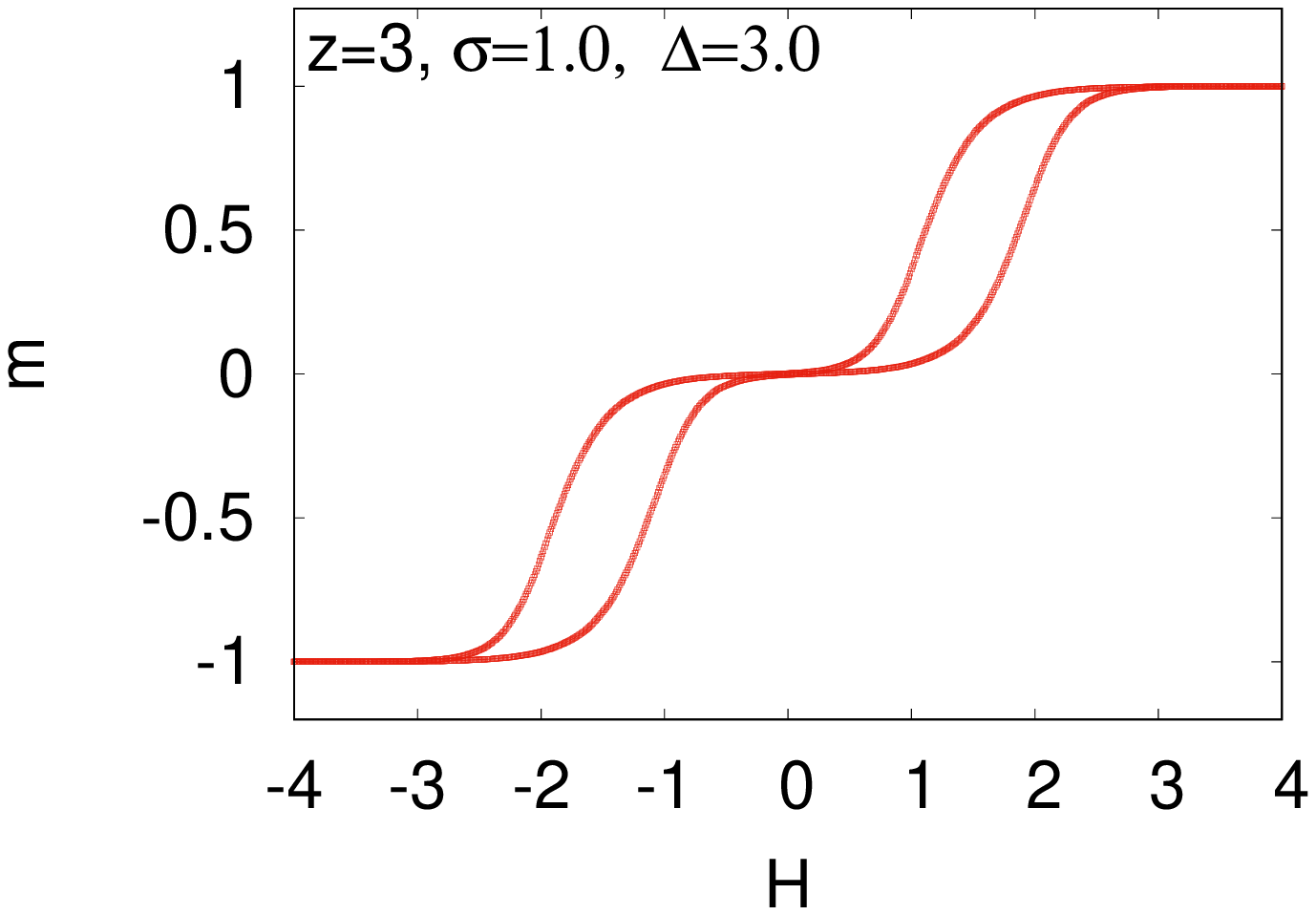}
\caption{(Appendix B) The magnetization for different values of the parameters ($\Delta = -1, 1, 3$ and $\sigma = 0.15, 0.5, 1$) for $z = 3$ are shown above.}
\label{B_z3}
\end{figure*}

\begin{figure*}[h!]
\includegraphics[width=0.65\columnwidth]{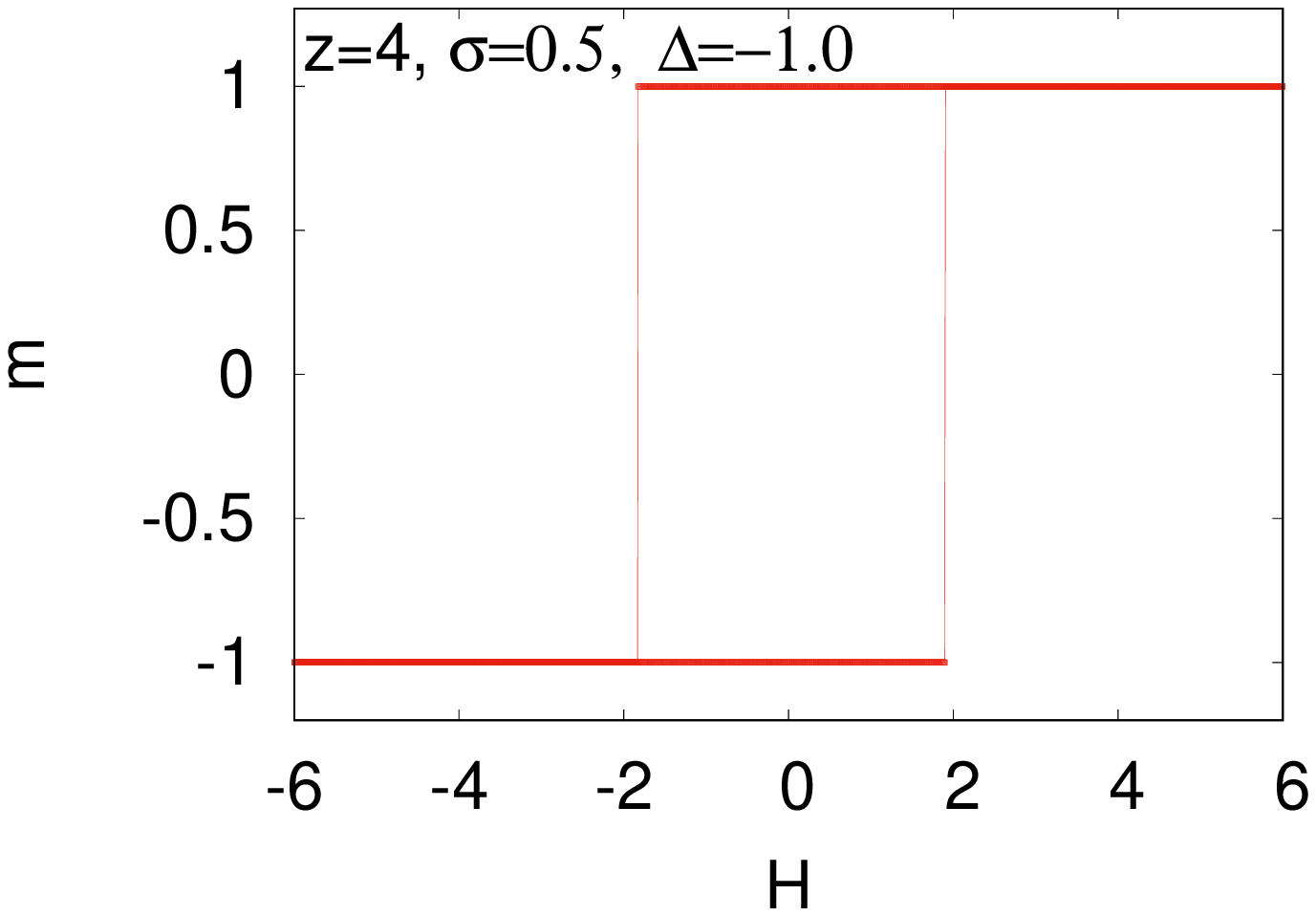}
\includegraphics[width=0.65\columnwidth]{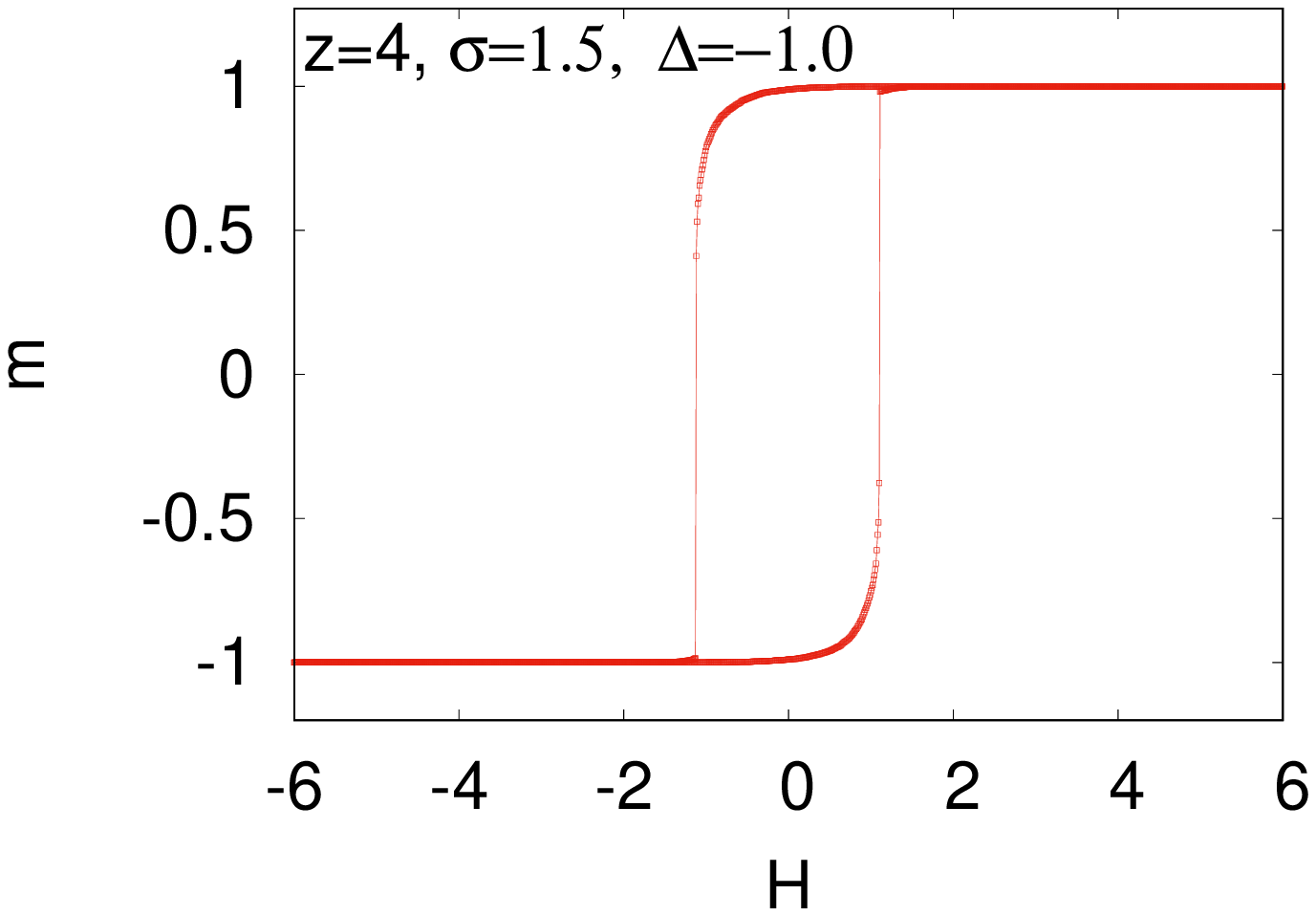}
\includegraphics[width=0.65\columnwidth]{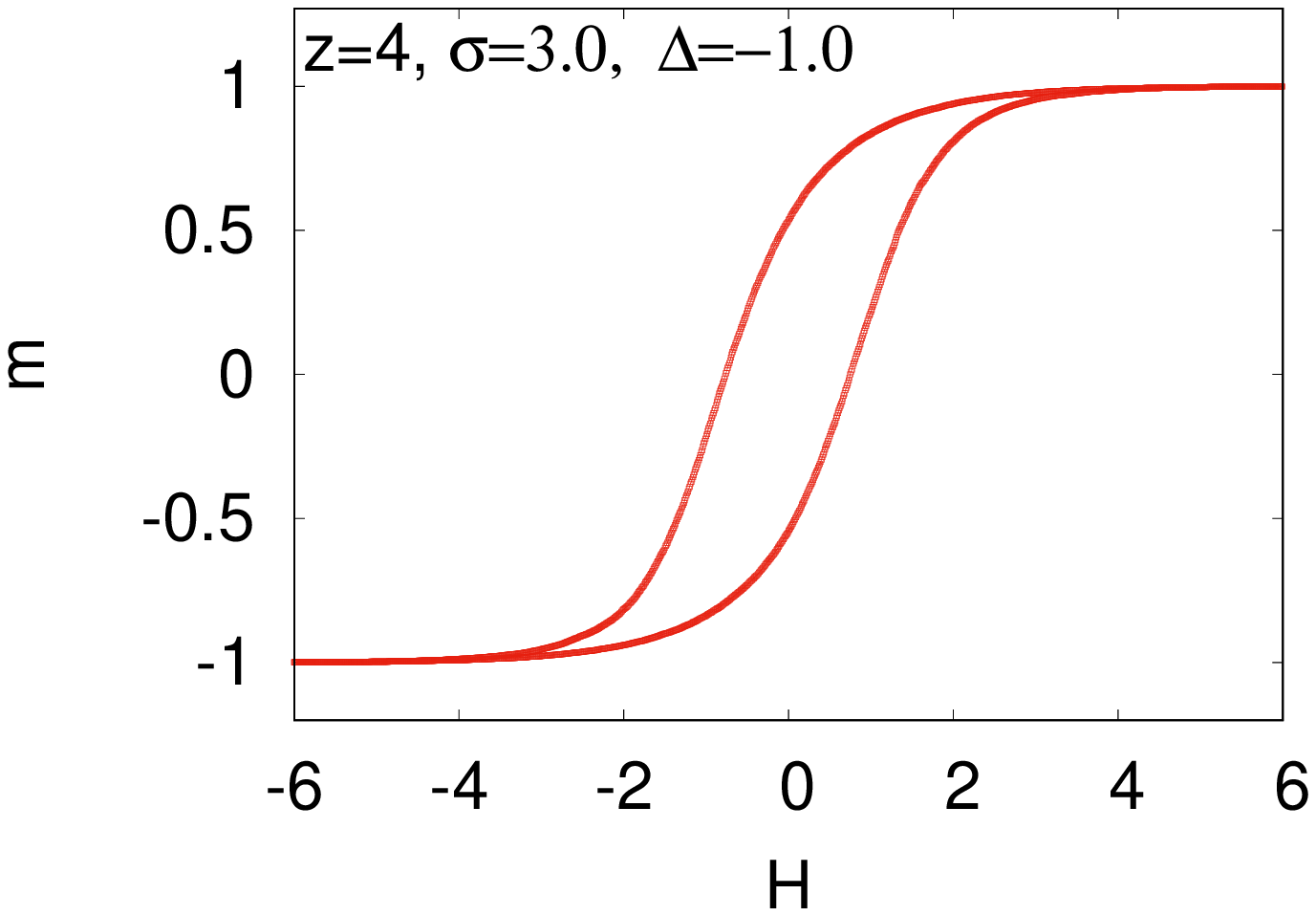}
\includegraphics[width=0.65\columnwidth]{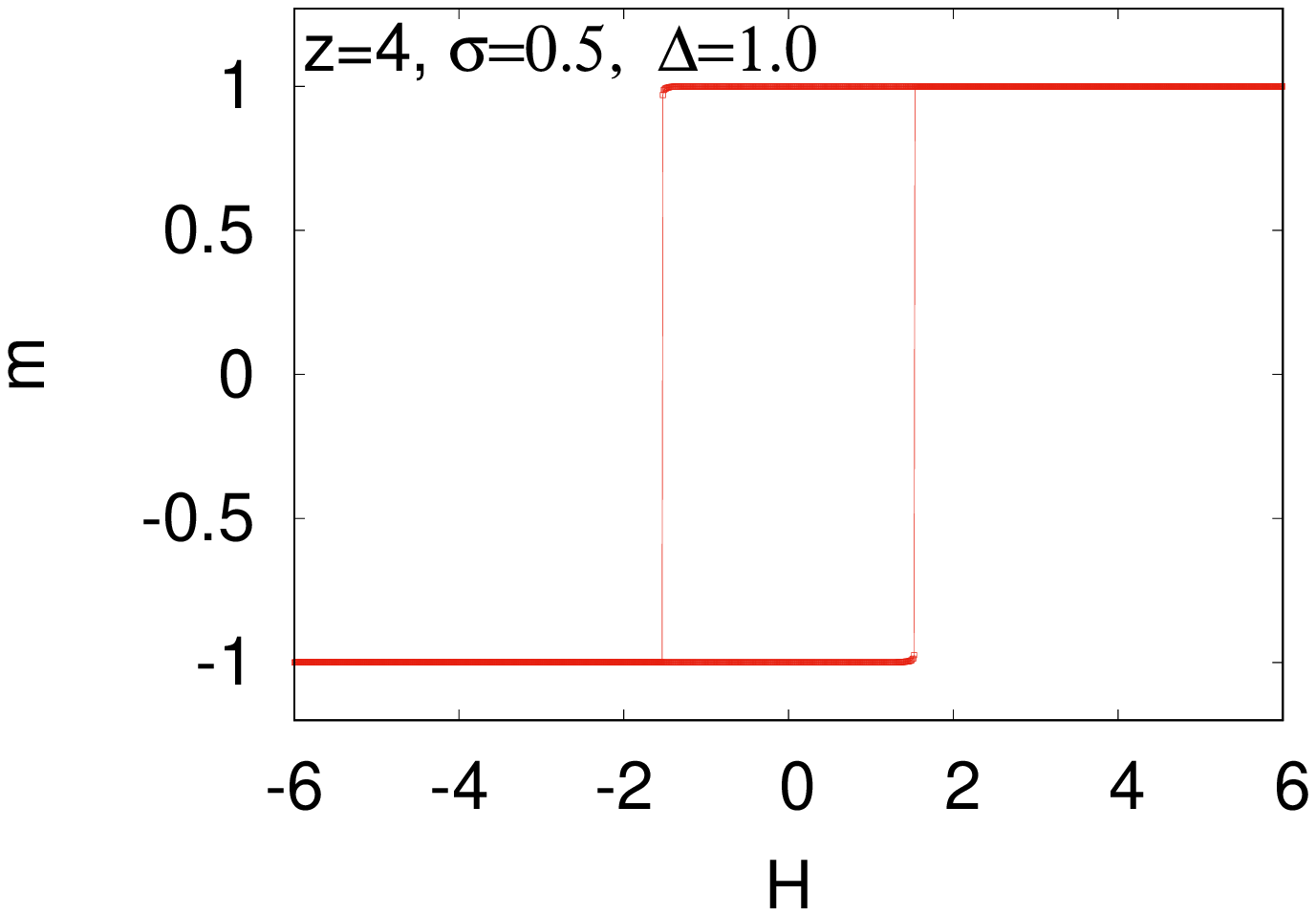}
\includegraphics[width=0.65\columnwidth]{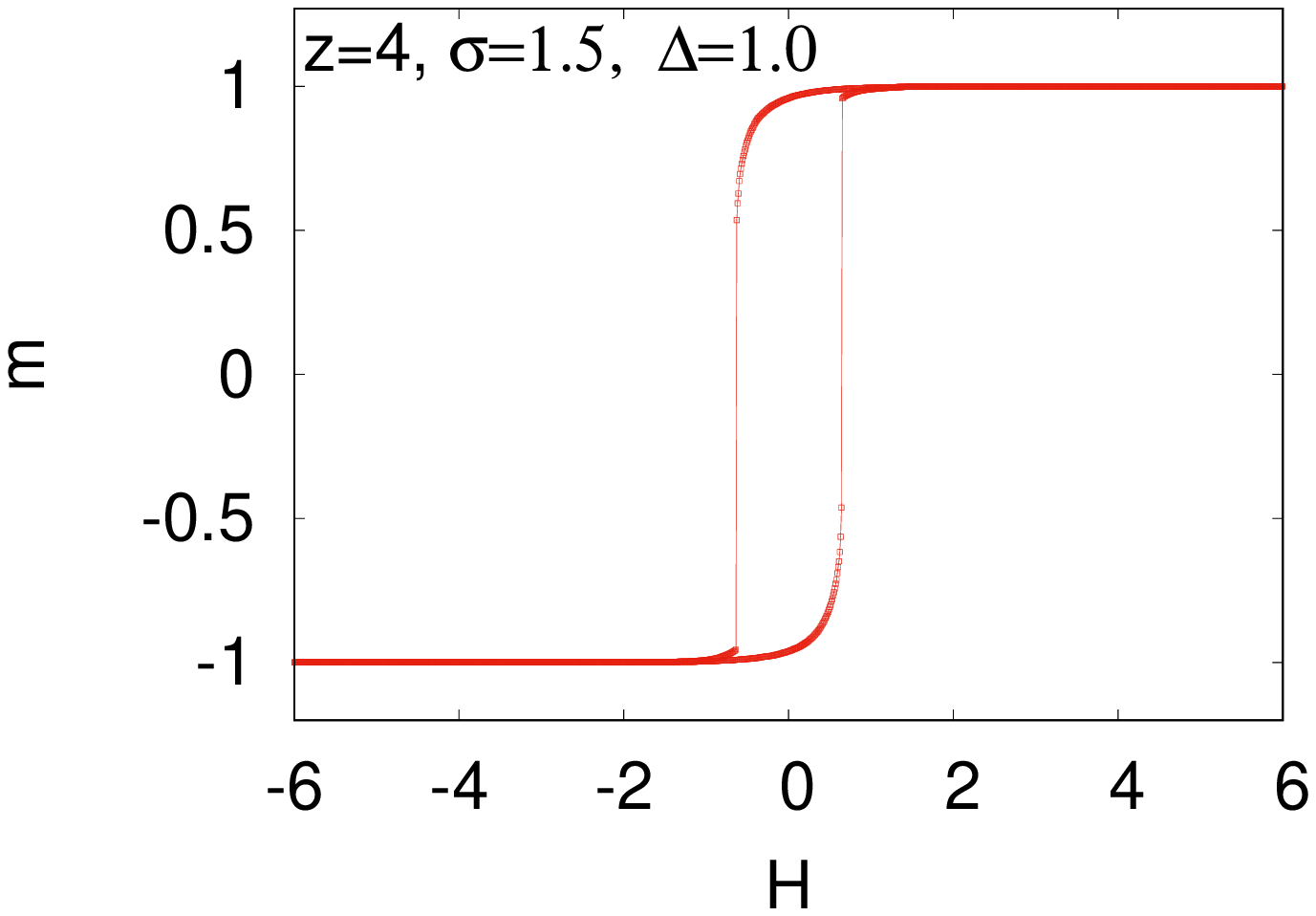}
\includegraphics[width=0.65\columnwidth]{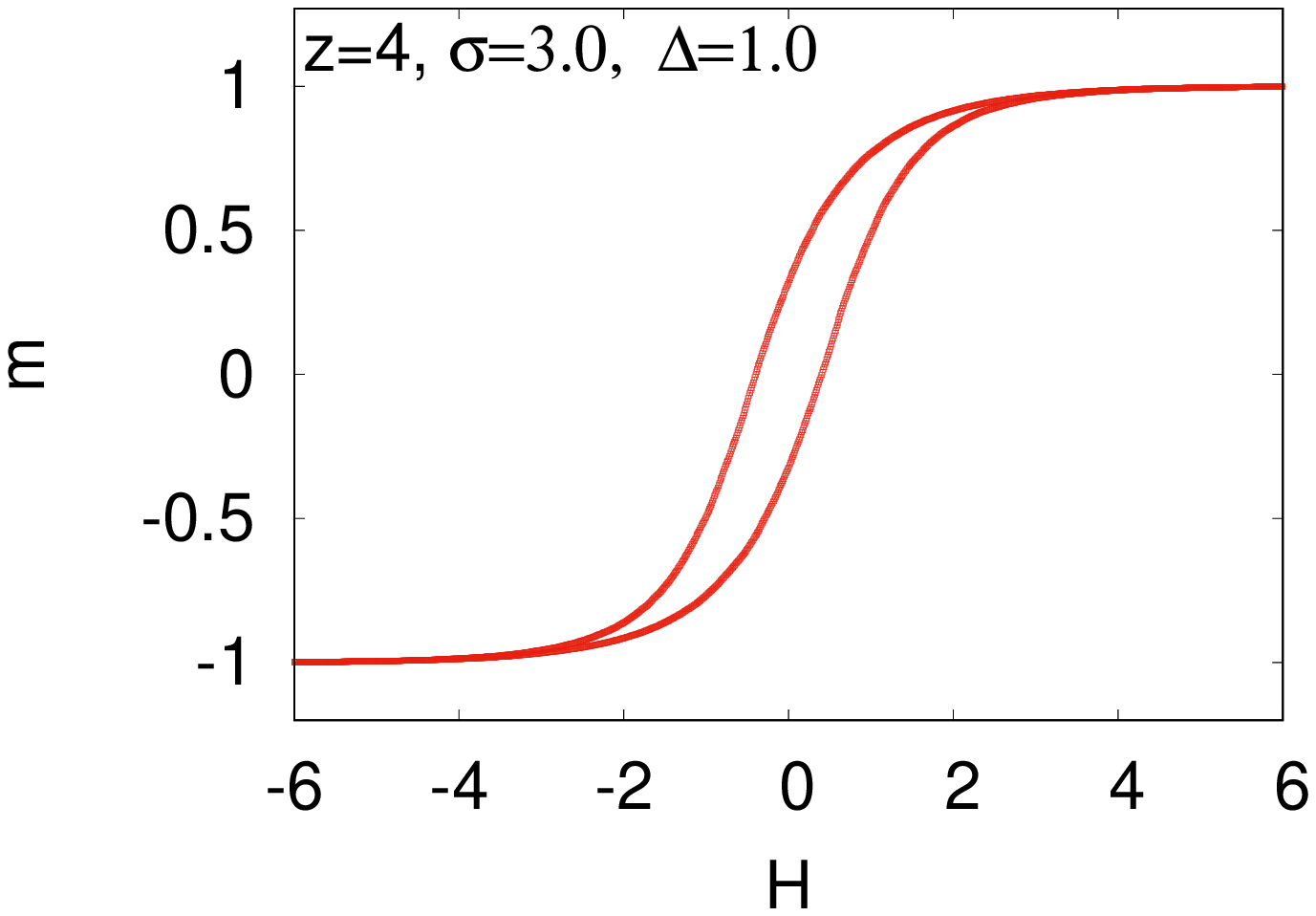}
\includegraphics[width=0.65\columnwidth]{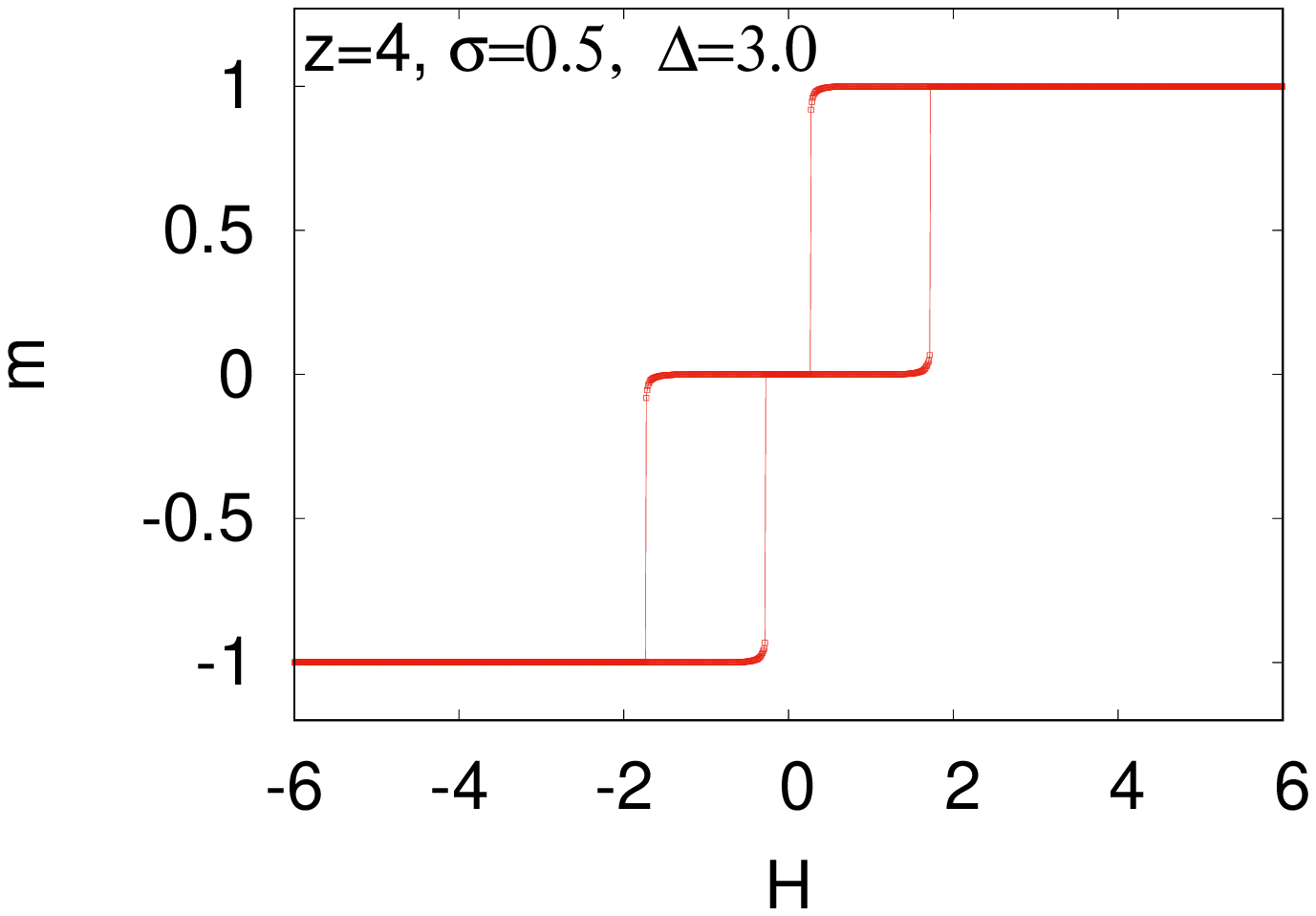}
\includegraphics[width=0.65\columnwidth]{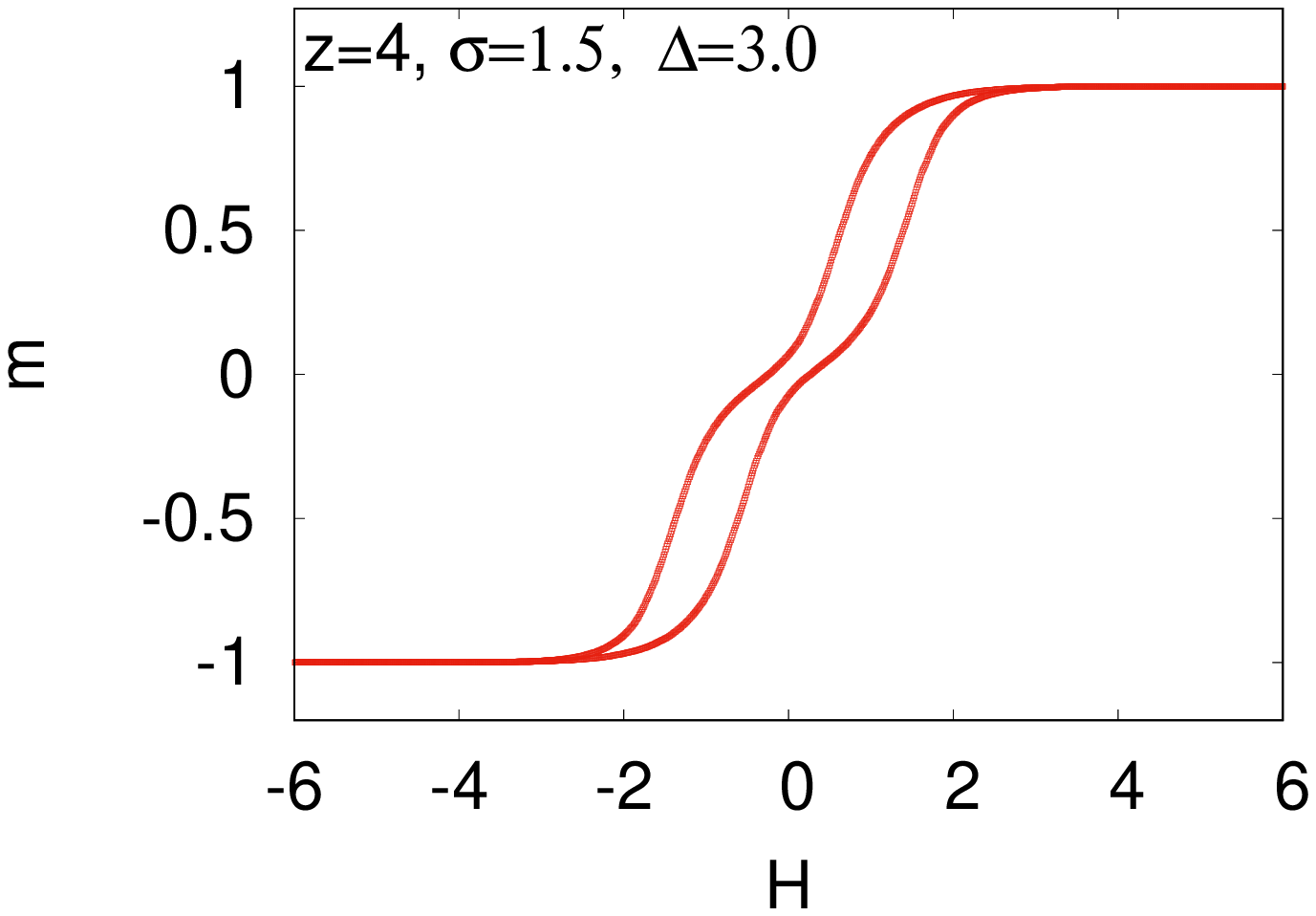}
\includegraphics[width=0.65\columnwidth]{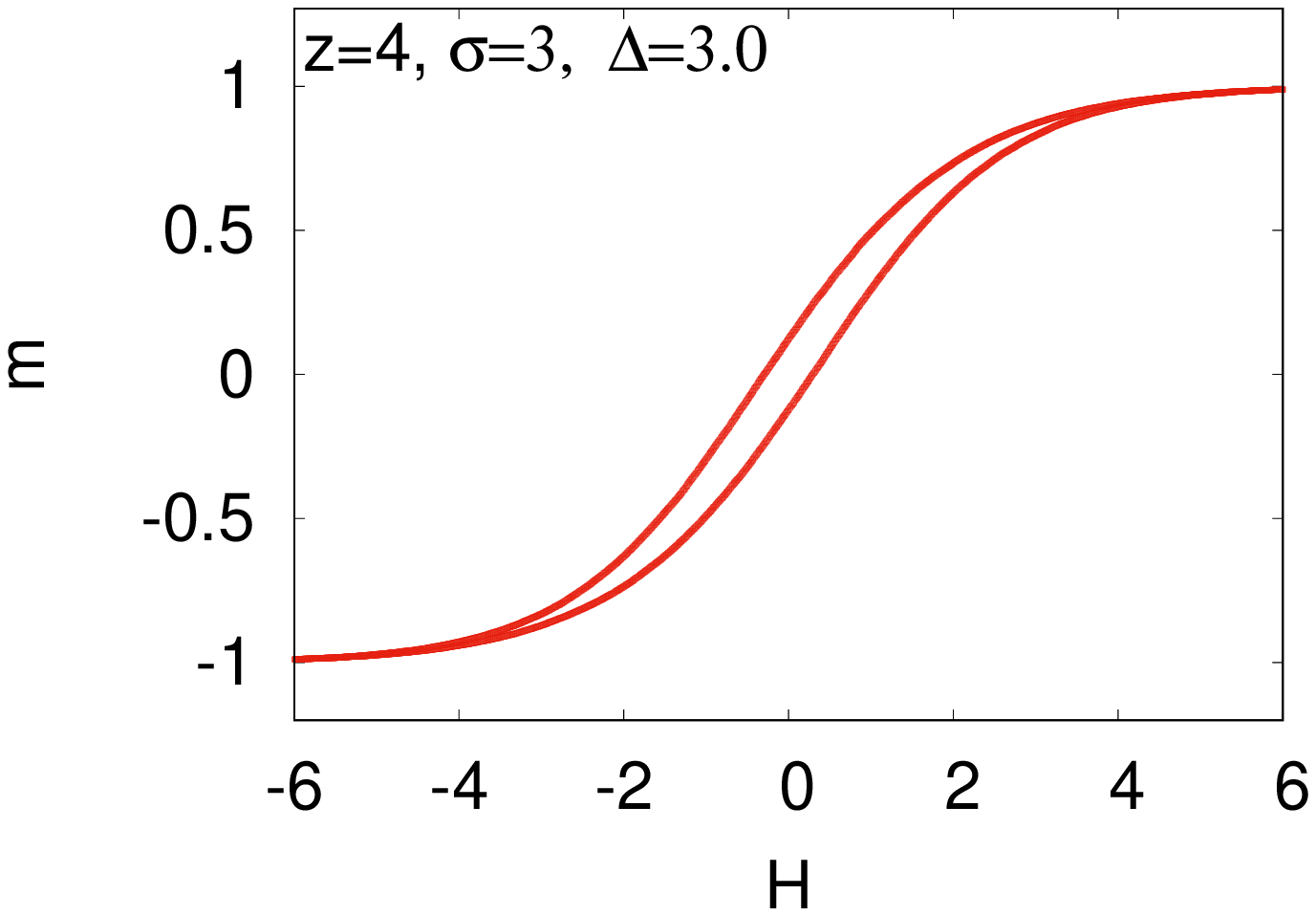}
\caption{(Appendix B) The magnetization for different values of the parameters ($\Delta = -1, 1, 3$ and $\sigma = 0.5, 1.5, 3$) for $z = 4$ are shown above.} 
\label{B_z4}
\end{figure*}

We plot the magnetization plots of RFBCM for different values of the parameters ($\Delta$ and $\sigma$) for $z = 3$ (Fig. \ref{B_z3}) and $z = 4$ (FIG. \ref{B_z4}).
\section{Comparison of the magnetization from Bethe calculations with the simulations}
\label{appC}
\begin{figure*}[h!]
\includegraphics[width=1\columnwidth]{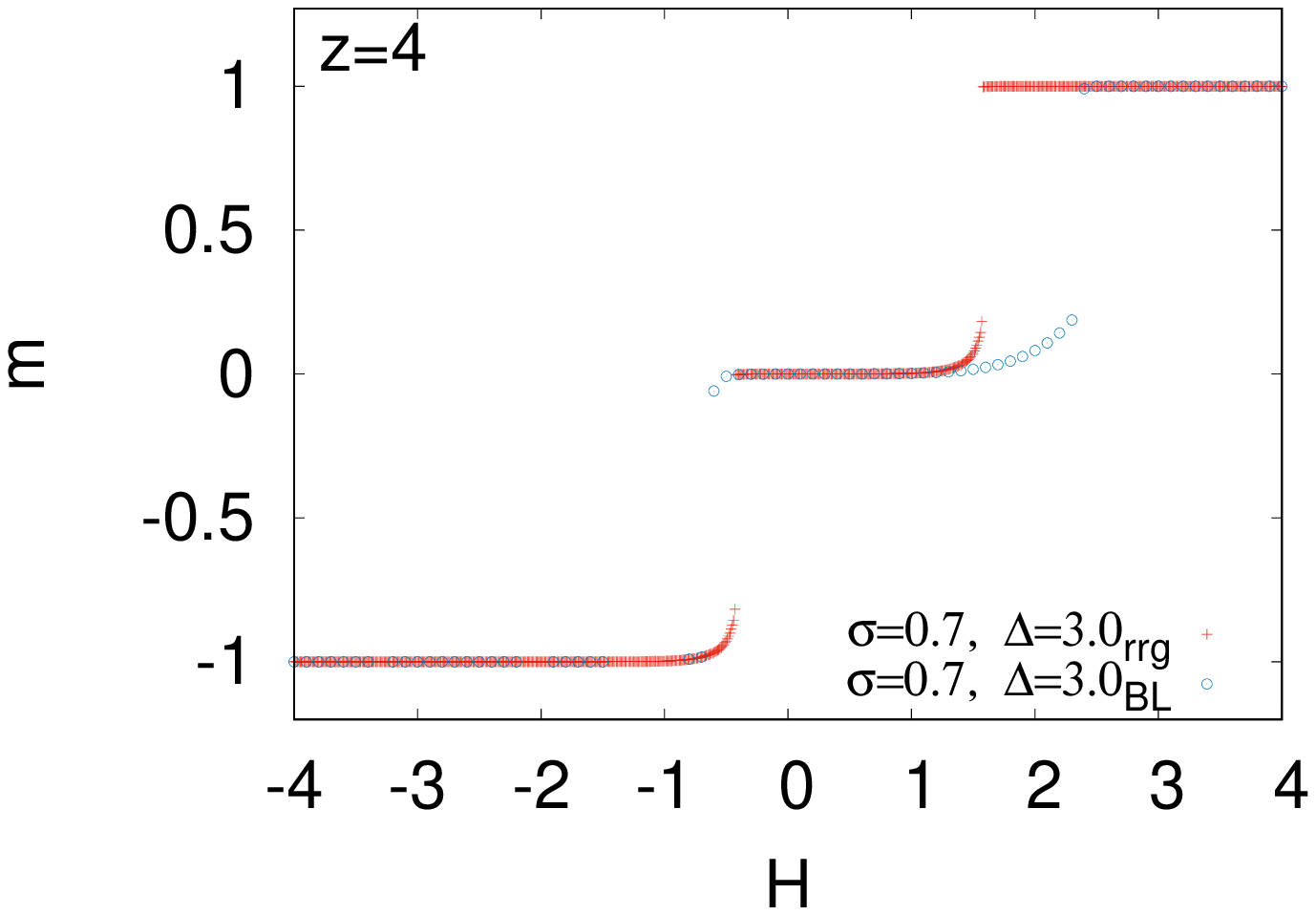}
\includegraphics[width=1\columnwidth]{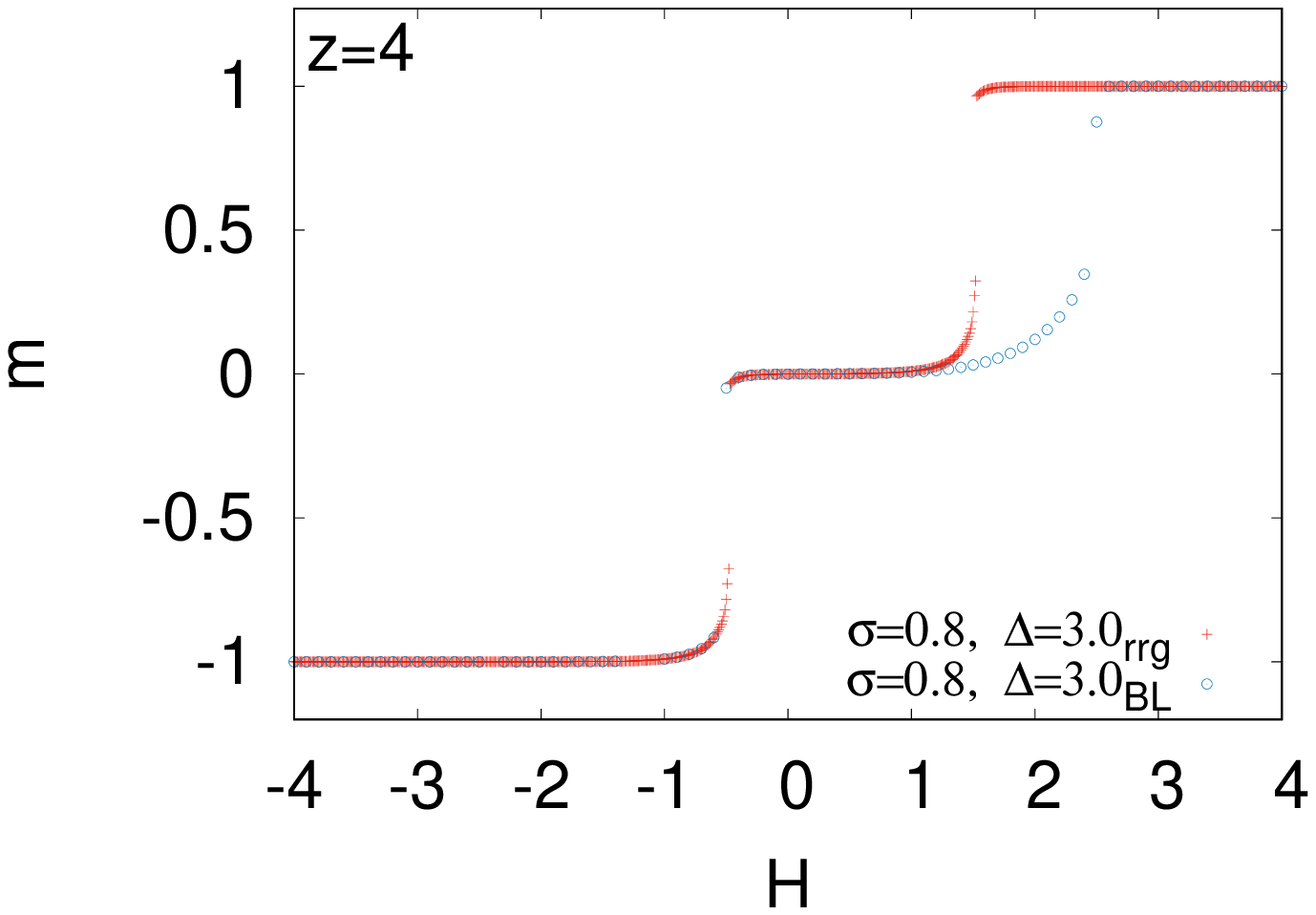}
\includegraphics[width=1\columnwidth]{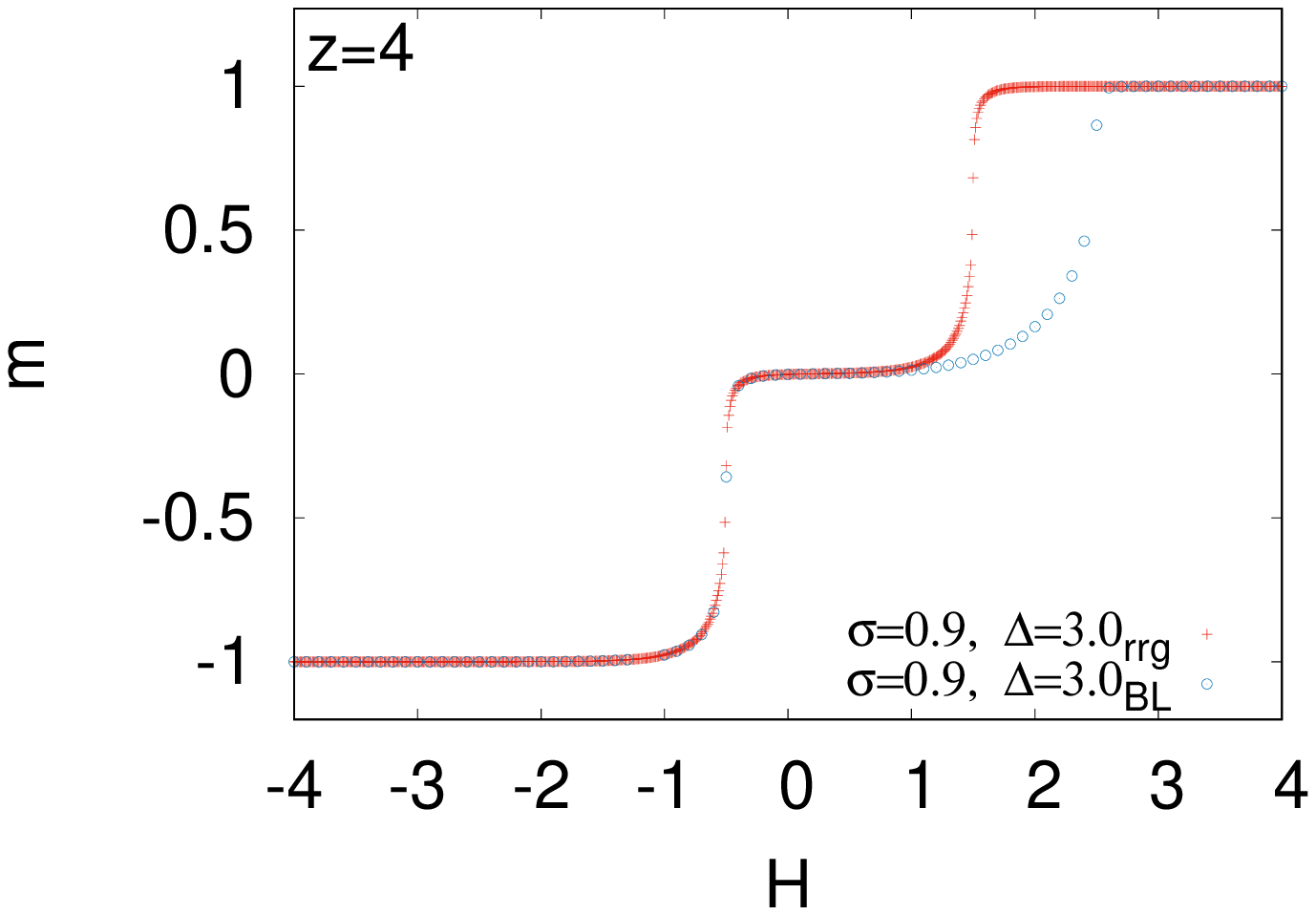}
\includegraphics[width=1\columnwidth]{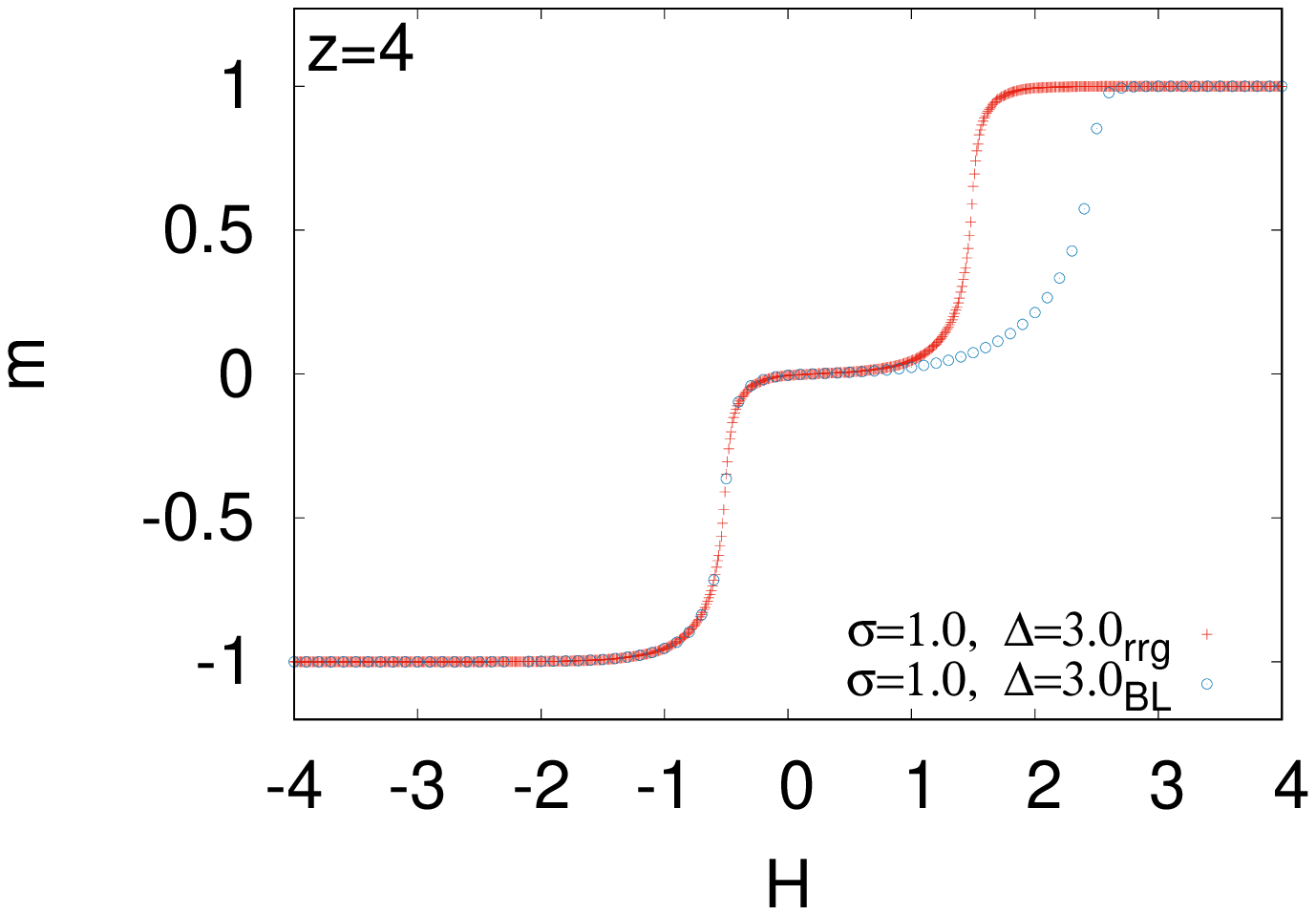}
\includegraphics[width=1\columnwidth]{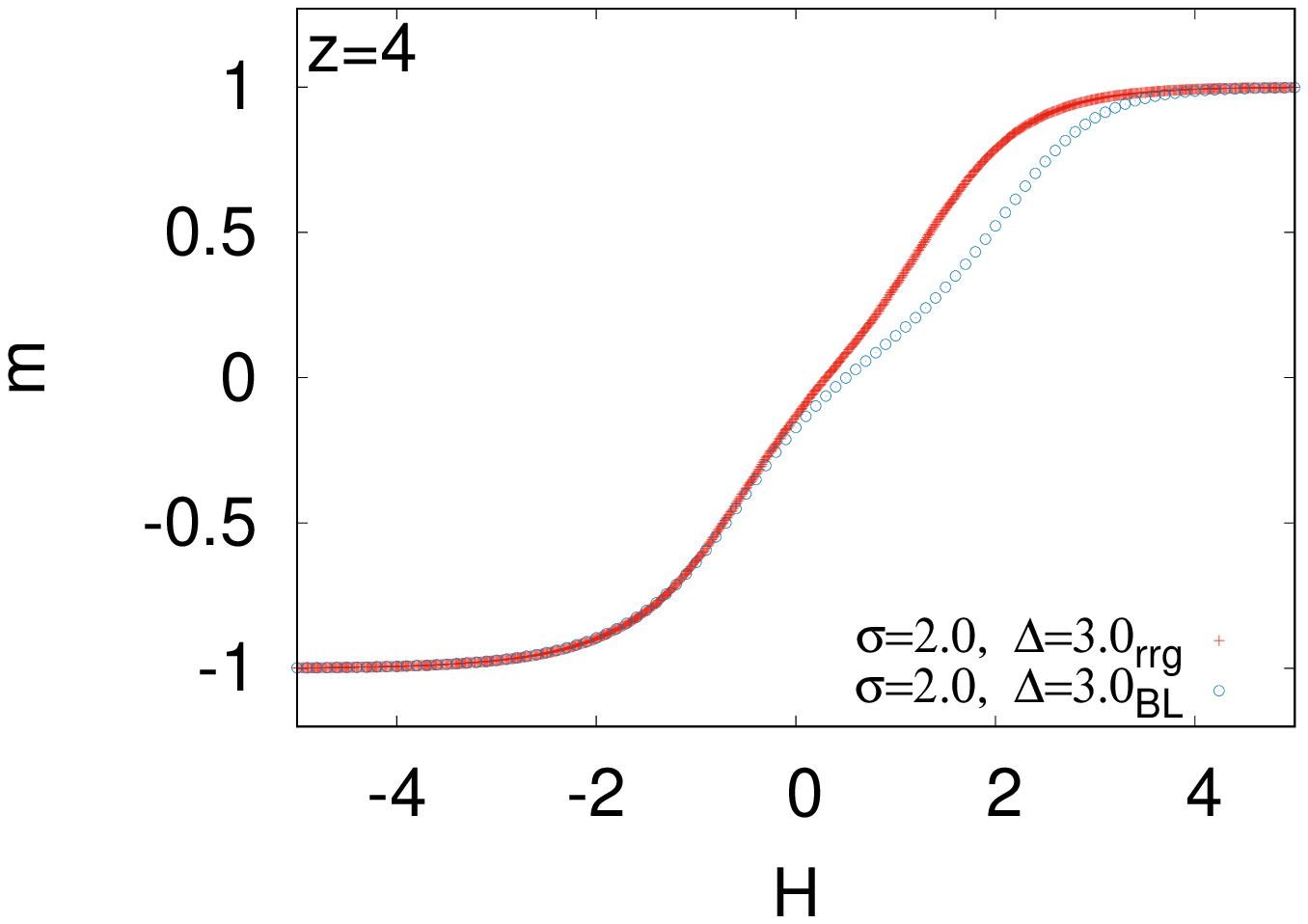}
\includegraphics[width=1\columnwidth]{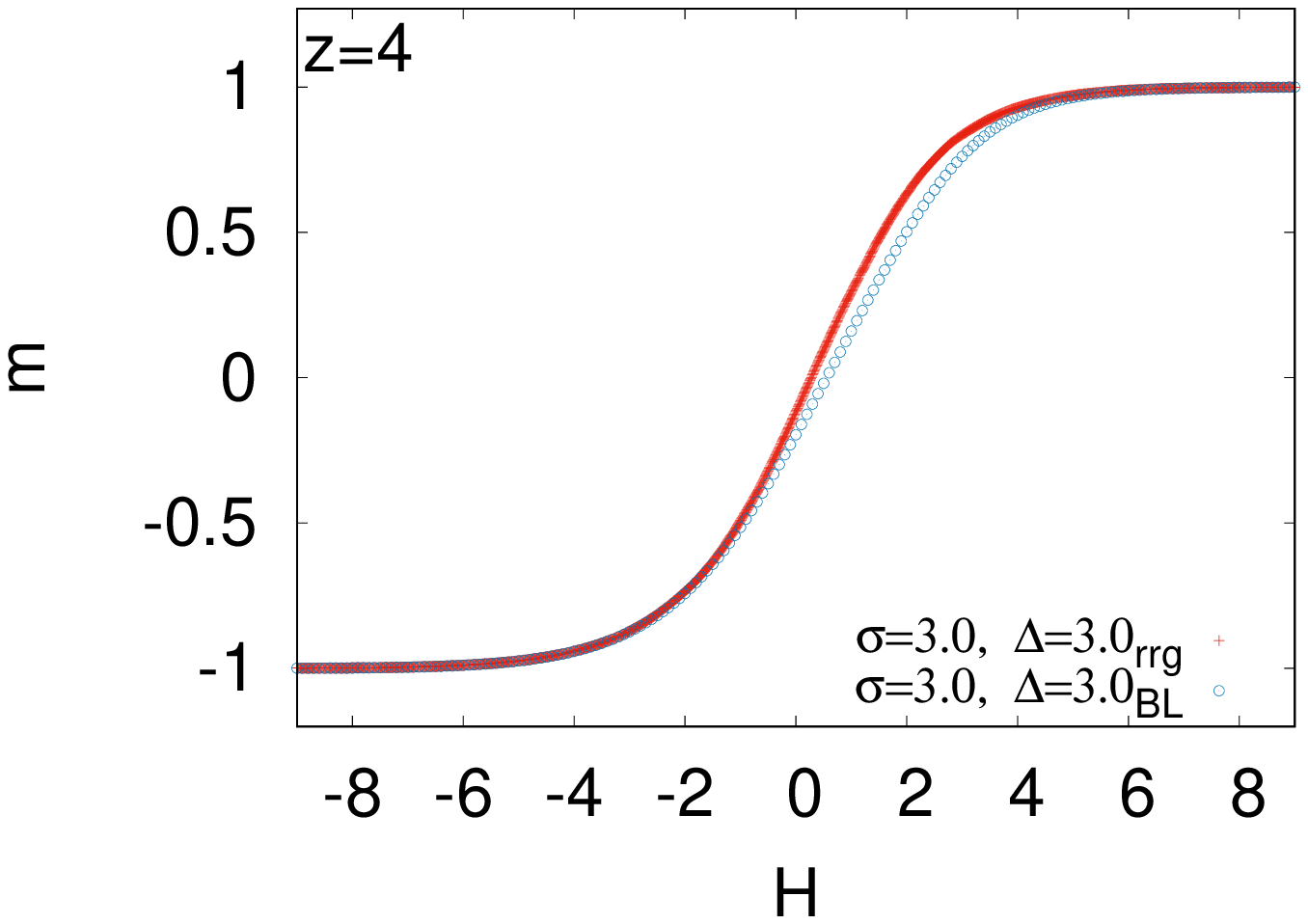}
\caption{(Appendix C) Magnetization from Bethe lattice (BL) calculations is compared with that from
simulations for forward magnetization curve on a random regular graph (rrg)  for $z = 4$ and $\Delta = 3$ for various values of, $\sigma = 0.7, 0.8, 0.9, 1, 2$ and $3$.}
\label{C}
\end{figure*}

In Fig.\ref{C}, we compare the magnetization plots obtained from Bethe lattice calculations with that obtained from the simulations for $z = 4$, $\Delta = 3$ for $\sigma= 0.7, 0.8, 0.9, 1, 2$ and $3$. In the negative $H$ region there is a good match between the simulations and the Bethe lattice calulcations. As expected, the mismatch increases with increasing $\sigma$.

{}


\begin{thebibliography}{}
\bibitem{footnote} The dynamics is considered abelian if the order in which unstable spins are relaxed does not matter.
\end{thebibliography}
\end{document}